\documentclass[letterpaper]{JHEP3}
\usepackage{epsfig}

\newcommand{\roughly}[1]{\mathrel{\raise.3ex\hbox{$#1$\kern-0.85em
\lower1ex\hbox{$\sim$}}}}

\newcommand{\lsim}{\roughly<}
\newcommand{\gsim}{\roughly>}

\def\pd{\partial}

\def\cA{{\cal A}}

\def\cC{{\cal C}}

\def\cF{{\cal F}}

\def\cL{{\cal L}}

\def\cO{{\cal O}}
\def\cQ{{\cal Q}}
\def\cR{{\cal R}}

\def\cU{{\cal U}}
\def\cV{{\cal V}}
\def\cW{{\cal W}}

\def\exd{{\hbox{d}}}

\def\ba{\begin{eqnarray}}
\def\ea{\end{eqnarray}}
\def\be{\begin{equation}}
\def\ee{\end{equation}}

\def\ssM{{\scriptscriptstyle M}}
\def\ssN{{\scriptscriptstyle N}}
\def\ssP{{\scriptscriptstyle P}}
\def\ssQ{{\scriptscriptstyle Q}}
\def\ssR{{\scriptscriptstyle R}}

\def\KK{{\scriptscriptstyle KK}}
\def\JF{{\scriptscriptstyle JF}}
\def\EF{{\scriptscriptstyle EF}}
\def\QCD{{\scriptscriptstyle QCD}}

\def\nn{\nonumber}

\def\({\left(}
\def\){\right)}

\def\pref#1{(\ref{#1})}

\title{Technically Natural Cosmological Constant\\
From Supersymmetric 6D Brane Backreaction}

\author{
C.P. Burgess${}^{1,2}$ and Leo van Nierop${}^1$ \\
$^1$Department of Physics and Astronomy, McMaster University, Hamilton ON, Canada\\
$^2$Perimeter Institute for Theoretical Physics, Waterloo ON, Canada\\
}

\maketitle

\abstract {We provide an explicit example of a higher-dimensional model describing a non-supersymmetric spectrum of 4D particles of mass $M$, whose 4D geometry -- {\em including loop effects} -- has a curvature that is of order $R \sim m_\KK^4/M_p^2$, where $m_\KK$ is the extra-dimensional Kaluza-Klein scale and $M_p$ is the 4D Planck constant. $m_\KK$ is stabilized and can in particular satisfy $m_\KK \ll M$. The system consists of a (5+1)-dimensional model with a flux-stabilized supersymmetric bulk coupled to non-supersymmetric matter localized on a (3+1)-dimensional positive-tension brane. We use recent techniques for calculating how extra dimensions respond to changes in brane properties to show (at the classical level) that the extra-dimensional volume adjusts to ensure that the low-energy 4D geometry is exactly flat, independent of the value of the brane tensions. Its mechanism for doing so is the transfer of stabilizing flux between the bulk and the branes. The UV completion of the model can arise at scales much larger than $M$, allowing the calculation of quantum effects like the zero-point energy of very massive particles in the vacuum. We find that brane-localized loops do not affect the 4D curvature at all, but bulk loops can. These can be estimated on general grounds and we show that supersymmetry dictates that they generate curvatures that are generically of order $m_\KK^4/M_p^2$. For realistic applications this points to a world with two supersymmetric extra dimensions, with supersymmetry in the bulk broken at the sub-eV KK scale  --- as proposed in {\tt hep-th/0304256} --- requiring a 6D gravity scale somewhat higher than 10 TeV. Ordinary Standard Model particles are brane-localized and not at all supersymmetric (implying in particular no superpartners or the MSSM). We discuss how the model evades various no-go theorems that would naively exclude it, and briefly outline several striking observational implications for tests of gravity and at the LHC.
}

\begin{document}

\section{Introduction}

The cosmological constant problem \cite{Wbgcc} remains an important conceptual obstacle to our understanding of the hierarchies of the physical world. The puzzle of why the electroweak scale is much smaller than the Planck (and possibly GUT) scale has motivated many proposals for what kinds of physics might lie at TeV energies --- supersymmetry, compositeness, extra dimensions and so on --- that have been famously used to motivate many choices made when designing the now-operational Large Hadron Collider (LHC). But the same reasoning applied at the much lower, sub-eV energies relevant to the scale of Dark Energy seems to fail to explain how the vacuum energy can gravitate as weakly as it appears to do.

Many have remarked that extra dimensions (and large ones) can help with the cosmological constant problem, because they break the connection between the energy density of a 4-dimensional vacuum (which we believe should be large), and the curvature of the visible universe (which we observe to be small) \cite{preSLED, multibranes, 5Dbackreaction, SLED1, SLEDrevs}. The problem in four dimensions is that the Einstein equations force these to be the same, since $\langle T_{\mu\nu} \rangle = \frac14 \, T \, g_{\mu\nu}$ for any Lorentz-invariant state, implying $R_{\mu\nu} = - 2 \pi G \, T \, g_{\mu\nu}$. But once there are more than four dimensions then we need not demand the vacuum be Lorentz-invariant in the higher dimensions. And a large energy density that is Lorentz invariant in the 4D sense (such as a brane tension), can curve the extra dimensions rather than the four dimensions that are Lorentz-invariant. In particular, explicit solutions to the higher-dimensional field equations with brane sources are known that have this property, at least when there are not too many extra dimensions \cite{preSLED, 5Dbackreaction, SLED1}.

But the existence of some choices for brane sources for which bulk solutions can be flat is not in itself a solution to the cosmological constant problem. What must be shown is that these choices are sufficiently stable against integrating out heavy fields, including the electron.

In this paper we provide an explicit example of an extra-dimensional model which we believe predicts a 4D curvature whose size is controlled by the Casimir energy of the extra dimensions, $R \simeq m_\KK^4/M_p^2$, where $m_\KK$ is the Kaluza-Klein (KK) scale and $M_p$ is the Planck scale. In particular, it can be much smaller than what would be expected from the scales $M \gg m_\KK$ of particle physics. We regard it as a realization of an earlier general proposal --- supersymmetric large extra dimensions (SLED) \cite{SLED1} --- wherein ordinary particles are localized on a space-filling (3+1)-dimensional codimension-two brane that sits within a (5+1)-dimensional bulk spacetime with two compact dimensions transverse to the brane. We take the bulk to be described by a particular 6D supergravity (chiral, gauged supergravity \cite{NS}), but we believe the underlying mechanism applies equally well to other 6D supergravities, and more generally to other low-codimension brane systems interacting through bulk supergravities, once their back-reaction onto the bulk is accurately included.

A proper description of the vacuum energy must include in particular the energetics that stabilize the extra dimensions, and an advantage of the particular model we study is that many features of the extra-dimensions are stabilized within the bulk (without reference to the branes) at the classical level by a simple flux compactification \cite{SSs}, leaving only the single flat direction that is guaranteed by the classical scale invariance of the bulk supergravity.

It has been known for some time that brane back-reaction can lift this last flat direction \cite{uvcaps, TNc2B, BBvN, susybranes}, and what is new about our contribution here is to show that there is a simple choice for the brane-bulk couplings that can fix this last flat direction without generating an on-brane curvature, assuming we work only to within the classical approximation in the bulk (more about quantum corrections below). Two brane properties are required: $(i)$ the absence of a direct brane coupling to the bulk dilaton (a scalar superpartner of the graviton in six dimensions); and $(ii)$ a Maxwell-brane coupling that allows one of the branes to carry a localized amount of the bulk-stabilizing flux.

What is remarkable is that these properties are unchanged under arbitrary loops of the on-brane fields, including in particular loops of all ordinary particles of everyday experience (which we assume to be localized on one of the branes). This is possible because property $(i)$ --- the {\em absence} of a coupling to a bulk field --- is automatically preserved by brane loops if it is true at the classical level. Loops of ordinary particles also cannot alter the brane-localized flux coupling required by property $(ii)$ if the brane-bound particles do not couple to the bulk flux field. In the model explored below we assume the brane-flux coupling occurs on a different brane from that on which all brane particles are localized.

The requirement for brane-localized flux --- property $(ii)$ above --- turns out to be the new crucial ingredient, since it is the possibility of being able to localize some bulk fluxes onto the brane that allows the system to respond with little energy to changes of brane tension \cite{susybranes}. It is also this brane-localized flux that is responsible for stabilizing the remaining flat direction in the bulk, which it can do because the bulk flux field couples to the 6D scalar dilaton that parameterizes this flat direction.

We lay out our arguments in the following way. First, the remainder of this section carefully defines the notion of `technical naturalness,' whose absence is the essence of the cosmological constant problem. We do so because we believe that a resolution of this problem ultimately points towards a world with two supersymmetric extra dimensions at sub-eV scales. Although this seems an extreme possibility, there seem to be no alternatives short of abandoning technical naturalness altogether. In the words of Sherlock Holmes \cite{SH} ``...when you have eliminated the impossible, whatever remains, however improbable, must be the truth.''

\S2\ then describes in detail the simple bulk and brane systems on whose properties our proposal rests. In particular this section describes the exact classical solutions that capture the back-reaction of the branes to the bulk and which govern the geometry of both the on-brane and off-brane dimensions. The size of quantum corrections is the topic of \S3, which studies both the implications of loops of on-brane and bulk modes. This section argues why loops of brane-localized fields do not change the conclusions of \S2\ at all, and why the leading contributions come from bulk loops only. The contributions of massless and massive fields in the bulk are contrasted, and both are argued only to generate contributions to the low-energy 4D scalar potential that are of order $m_\KK^4$. Our conclusions are summarized in \S4, including a qualitative discussion of why both extra dimensions and supersymmetry are required. This section closes with a brief summary of what is known about the potentially rich observational signatures that are implied by the present framework, together with a summary of issues needing further study. Three appendices deal with technical issues about localizing flux on branes; calculate the low-energy 4D potential for arbitrary small perturbations to the brane actions; and examine two common objections to the possibility of using extra dimensions to help with the cosmological constant problem.

\subsection{Technical naturalness (without cutoffs)}

Notions of `technical naturalness' are central to our motivation, so we pause here to state these carefully. We believe our discussion largely keeps to the party line, though we make an effort not to cast the issues in terms of cutoffs in divergent integrals. Those familiar with the issues should feel free to skip this section entirely.

An understanding of hierarchies of scale, like the electroweak hierarchy or the cosmological constant problems, comes in two parts. The first part asks why the hierarchy of scales exists in the first place in the fundamental theory at very small distances. Because this question is sensitive to physics at the fundamental scale --- possibly the string scale or some other quantum gravity scale --- it might not be answered until we ultimately understand this fundamental theory in detail. The second part of the understanding asks why the hierarchy is stable when various massive states are integrated out to produce one of the effective theories that describes the implications of the fundamental theory at the low energies we can observe.

Technical naturalness is addressed at this second part of the problem, since the low-energy effective theory is not unique (depending as it does on the energy range that is of interest for a particular low-energy observer). Yet it is implicit in our understanding of physics that a large hierarchy can be understood equally well in {\em any} of the effective theories for which we choose to ask the question.

For example, a large hierarchy that is well-understood is the small size of the nucleus, $\ell_n$, relative to the size of an atom, $a_0$. Within the standard model the small ratio $\ell_n/a_0 \simeq 10^{-5}$ is understood as being a consequence of two other experimental facts: the electromagnetic coupling constant is weak: $\alpha = e^2/4\pi \simeq 10^{-2}$; and the electron is light compared with the QCD scale, $m_e / \Lambda_{\QCD} \simeq 10^{-3}$. The small size of the nucleus is then a consequence of the relations $a_0^{-1} \simeq \alpha m_e$ and $\ell_n^{-1} \simeq \Lambda_\QCD$.

But the same question might again be asked within a lower-energy effective theory below the QCD scale, say within the quantum electrodynamics of electrons, protons and neutrons. In this theory the small ratio of observables, $\ell_n/a_0$, is instead understood as a consequence of the small size of $\alpha$ in this theory, together with the electron being much lighter than the proton: $m_e \ll m_p$. The process of integrating out the quarks and gluons to give the proton and neutron (or integrating out the muon or other particles) does not fundamentally change the way we think about nuclei being small, and this is what it means\footnote{As originally formulated \cite{TNdef}, a small ratio is said to be technically natural if a new symmetry emerges when the ratio goes to zero. This is a particularly important way of ensuring technical naturalness in the way we define it here.} for this hierarchy to be `technically natural.'

Contrast this with our understanding of the small ratio between the observed Dark Energy density, $\rho_{\rm vac}$, and the electron mass (say): $\rho_{\rm vac}/m_e^4 \simeq 10^{-36}$. Consider the low-energy theory well below the electron mass, for which the fundamental particles might be taken to be photons, neutrinos and gravitons. For this theory $\rho$ is given by the cosmological constant that appears in the low-energy effective action, plus the loop contributions of these very low-mass particles:
\be
 \rho_{\rm vac} \simeq \lambda_{\rm le} + \hbox{low-energy loops}\,,
 \quad \hbox{where} \quad
 \cL_{\rm eff} = - \sqrt{-g} \; \Bigl( \lambda_{\rm le} + \cdots \Bigr) \,.
\ee

Compare this with the same calculation, performed in the effective theory defined above the electron mass, containing electrons in addition to the previously considered low-energy particles. There is an effective cosmological constant, $\lambda_{\rm he}$, also in this effective theory, whose value is related to $\lambda_{\rm le}$ by a matching condition that states that the physical quantity, $\rho$, should be the same in this theory as in the lower-energy effective theory. This implies that the renormalized value of $\lambda_{\rm he}$ in the effective theory above the electron mass is related to that below, $\lambda_{\rm le}$, by
\be \label{lambdashift}
 \lambda_{\rm le} \simeq \lambda_{\rm he} + \frac{k \, m_e^4}{16\pi^2} \,,
\ee
where $k$ is an order-unity number whose value is computed by evaluating an electron loop graph. This kind of shift, $\lambda_{\rm le} \to \lambda_{\rm he}$ occurs as we match across the electron threshold because the low-energy theory is obtained by integrating out the electron, meaning electrons are not present there to contribute to $\rho$ through loops. Eq.~\pref{lambdashift} expresses how $\lambda$ must change between the two theories to ensure that the low-energy theory `knows' about the contributions of virtual electrons to the vacuum energy.

Now comes the main point. Since all of the masses of particles in the very low-energy theory below the electron threshold are small, $\lambda_{\rm le}$ is of the same order of magnitude as is $\rho_{\rm vac}$. Consequently eq.~\pref{lambdashift} implies $\lambda_{\rm he}$ must be much larger than $\rho$ in the effective theory above the electron threshold. This nevertheless produces a small value for $\rho$ in this higher-energy theory because of a cancelation of roughly 36 decimal places between $\lambda_{\rm he}$ and an equally large electron loop, with the much smaller value, $\lambda_{\rm le}$, of the lower-energy theory emerging as the residue.

Instead of there being an understanding in {\em all} effective theories why $\lambda$ is smaller than $m_e^4$, the small size of $\lambda_{\rm le}$ in the very low-energy theory is understood as arising as an incredibly detailed cancelation between much larger quantities like $\lambda_{\rm he}$ and loops of the many much heavier particles the higher-energy theories contain. Of course, it is logically possible that this is the way nature works. But although we know about very many other hierarchies in nature (besides that between atoms and nuclei), so far as we know {\em none} of these are understood in this way.

It is a radical proposal that advocates that new hierarchies should be understood so very differently than those we've understood well in the past. A more scientifically conservative approach is instead to seek a technically natural understanding of poorly understood hierarchies like the electroweak hierarchy and the Dark Energy density. Of course this is a very tall order for the Dark Energy, since its very small size means that any successful approach must modify the properties of comparatively low-energy particles (like the electron).

\subsubsection*{Purging cutoffs}

Notice that the previous paragraphs are all formulated in terms of physical, or of renormalized, masses. Technical naturalness is often stated in a cutoff-dependent way, in terms of the absence of quadratic or quartic divergences when loop contributions are computed within a low-energy effective theory. We deliberately do not phrase things here in terms of cutoffs because we believe this can be a confusing way to express the physical issues at stake.

At face value quadratic dependence on a cutoff sounds like the same thing as sensitivity to heavy particle masses, because within a cutoff regularization the value of the cutoff very concretely specifies where the high-energy theory starts and a low-energy effective theory breaks down. Furthermore, quadratic or higher dependence on a cutoff indicates a strong sensitivity of a loop integral to the details of the unknown high-energy physics. However, from the point of view of a Wilsonian effective field theory, cutoffs are one of the few things we can be sure never enter into physical quantities, because they are an artefact of how a theorist decides to organize a calculation into a low- and high-energy contribution  \cite{WilsonEFT, eftreviews}. In particular, damage done by using a silly or inconvenient regularization, can be undone by appropriately renormalizing the resulting theory.

{}From this point of view the scale of the cutoff in the low-energy theory is really only a proxy for a bona fide mass of a state in the UV completion, and the presence of quadratic divergences really only provide a qualitative indication of when heavy masses can appear as an enhancement when integrating out a heavy particle. But in the end, the relation between cutoffs and heavy masses is not quantitative, and to properly decide whether heavy masses contribute significantly to an observable really requires knowledge of the UV completion that describes its properties, and cannot be decided purely within the low-energy theory. In general, the coefficients of quadratic divergences do not track those of heavy masses, and one can get burned by taking the correspondence between heavy masses and cutoffs too seriously \cite{uses}.

\section{Classical brane-bulk dynamics}

We start by summarizing the bulk-brane system of interest, which we choose closely following ref.~\cite{susybranes}. Since our results depend only on the dynamics of codimension-2 branes within higher-dimensional supergravity (together with the classical scale invariance these supergravities naturally enjoy), we believe our results not to be limited to the specific six-dimensional supergravity we examine here in detail.

\subsection{The bulk system}

We take chiral gauged supergravity in six dimensions \cite{NS} to govern the bulk physics, to which we couple two space-filling, positive-tension branes that strongly break supersymmetry. The bulk fields whose dynamics we follow in detail are the metric, $g_{\ssM\ssN}$; a flux-stabilizing bulk Maxwell gauge potential, $\cA_\ssM$; and the 6D scalar dilaton, $\phi$.

\subsubsection*{Field equations}

The bulk bosonic action restricted to these fields is\footnote{We use a `mostly plus' metric and Weinberg's curvature conventions \cite{Wbg} (that differ from those of MTW \cite{MTW} only by an overall sign in the definition of the Riemann tensor).}
\be \label{BulkAction}
 S_\mathrm{bulk} = - \int \exd^6 x \sqrt{-g} \; \left\{ \frac1{2\kappa^2} \, g^{\ssM\ssN}
 \Bigl( \cR_{\ssM \ssN} + \pd_\ssM \phi \, \pd_\ssN \phi \Bigr)
 + \frac14 \, e^{-\phi} \cF_{\ssM\ssN} \cF^{\ssM\ssN}
 + \frac{2 \, g_\ssR^2}{\kappa^4} \, e^\phi \right\} \,,
\ee
where $\cF = \exd \cA$ denotes the  gauge potential's field strength, and $\kappa$ and $g_\ssR$ are, respectively, the dimensionful coupling constants for gravity and for a specific $U_\ssR(1)$ symmetry of the supersymmetry algebra. The full gauged supergravity has more bosonic fields than this, but the rest can be set to zero consistent with their equations of motion. The background gauge field, $\cA_\ssM$, need not be the one that gauges the $U_\ssR(1)$ symmetry so its gauge coupling, $g$, need not equal $g_\ssR$.

The equations of motion from this action are the (trace reversed) Einstein equations
\be \label{BulkEinsteinEq}
 \cR_{\ssM\ssN} + \partial_\ssM \phi \, \partial_\ssN \phi
  + \kappa^2  e^{-\phi}\cF_{\ssM \ssP} {\cF_\ssN}^\ssP
  - \left( \frac{\kappa^2}{8} \,
  e^{-\phi} \cF_{\ssP\ssQ} \cF^{\ssP \ssQ}
  - \frac{g_\ssR^2}{\kappa^2} \, e^\phi  \right)
  g_{\ssM\ssN} = 0 \,,
\ee
the Maxwell equation
\be \label{BulkMaxwellEq}
 \nabla_\ssM (e^{-\phi}\cF^{\ssM \ssN}) = 0 \,,
\ee
and the dilaton equation
\be \label{BulkDilatonEq}
 \Box \phi - \frac{2 \, g_\ssR^2 }{\kappa^2} \, e^\phi  + \frac{\kappa^2}4 \, e^{-\phi}
 \cF_{\ssM\ssN} \cF^{\ssM\ssN} = 0 \,.
\ee

These field equations enjoy the exact classical symmetry
\be \label{classscaleinv}
 g_{\ssM \ssN} \to \zeta \, g_{\ssM \ssN} \quad \hbox{and} \quad
 e^{-\phi} \to \zeta \, e^{-\phi} \,,
\ee
with $\cA_\ssM \to \cA_\ssM$. This ensures the theory has three important properties:
\begin{itemize}
\item It ensures any nonsingular solution is always part of a one-parameter family of solutions that are exactly degenerate (within the classical approximation);
\item It ensures that there exists a Weyl rescaling, $\check g_{\ssM \ssN}
    = e^\phi \, g_{\ssM \ssN}$, for which $\phi$ appears undifferentiated in the bulk action only as an overall factor. That is,
    \be \label{phioutfront}
     S_\mathrm{bulk} = - \int \exd^6x \sqrt{-\check g} \; e^{-2\phi}
     \cL(\check g_{\ssM \ssN}, \partial_\ssM \phi, F_{\ssM \ssN}) \,,
    \ee
    where $\cL$ only depends on derivatives of $\phi$. Eq.~\pref{phioutfront} is significant because it shows that the quantity $e^{2\phi}$ plays the role of $1/\hbar$, and so is the loop-counting parameter for the bulk part of the theory.
\item It ensures that once evaluated at {\em any} solution of the field equations --- {\em i.e.} eqs.~\pref{BulkEinsteinEq} through \pref{BulkDilatonEq} --- the action, eq.~\pref{BulkAction}, evaluates to a total derivative \cite{WBW6D},
    \be \label{BulkSclass}
     \Bigl. S_\mathrm{bulk} \Bigr|_{\rm soln} = \frac1{2\kappa^2}
     \int \exd^6 x \sqrt{-g} \; \Box \phi \,.
    \ee
\end{itemize}

\subsection{Brane properties}

We focus on configurations involving two space-filling (3+1)-dimensional branes, whose coupling to the bulk fields we take to be given by the leading terms in a derivative expansion:
\be \label{BraneFluxCoupling}
 S_\mathrm{branes} = - \sum_{b} \int d^4x \sqrt{-g_4} \;
 \left[ T_b(\phi) - \frac12 \, \Phi_b(\phi) \, \epsilon^{mn} \cF_{mn}
  + \cdots \right] \,,
\ee
where the ellipses indicate terms involving two derivatives or more.\footnote{Notice that we normalize the quantity $\Phi_b$ slightly differently than in ref.~\cite{susybranes}.} In general the coupling functions $T_b$ and $\Phi_b$ can depend on $\phi$, as well as any fields localized on the branes (which we denote collectively by $\psi$). If $T_b$ is independent of $\phi$ and $\Phi_b \propto e^{-\phi}$, then the brane action transforms under the classical scaling symmetry, eq.~\pref{classscaleinv}, in the same way as does the bulk action, ensuring that the brane couplings do not break the classical bulk scale invariance.

The parameter $T_b$ represents the tension of the brane, and our conventions are such that $T_b > 0$ corresponds to positive tension. The parameter $\Phi_b$ corresponds physically \cite{susybranes} to the amount of magnetic flux that is localized on the source branes (see eq.~\pref{fluxquantzn} below). When $T_b$ drops out of the low-energy energetics -- as is the case below -- keeping nominally subdominant terms in the derivative expansion like the magnetic coupling to the brane becomes important \cite{SLED1}.

Let us now specify the properties of the two source branes in more detail. First is the observer's brane, $S_o$, on which all ordinary particles are imagined to reside,
\be \label{observerbrane}
 T_o = \tau_o + g^{\mu\nu} \partial_\mu \psi^* \partial_\nu \psi + M^2 \psi^* \psi + \cdots \,,
\ee
and on which there is no flux,\footnote{Since our conclusions depend only on the total brane flux, $\Phi_o + \Phi_f$, the vanishing of $\Phi_o$ is not necessary.} $\Phi_o = 0$. $\psi$ here could represent the Higgs boson, but more broadly is meant as a proxy for all of the fields of the Standard Model. The goal of later sections is to show that the on-brane curvature can be made systematically small compared with $M^4/M_p^2$ in a technically natural way.

Second is what we call the `flux' brane, on which no fields are localized and for which
\be \label{fluxbrane}
 T_f = \tau_f  \quad \hbox{and} \quad
 \Phi_f = \mu \,.
\ee
Here $\tau_o$, $\tau_f$, $\mu$ and $M$ are dimensionful parameters that define the energy scales of the system. Although the validity of semiclassical reasoning requires quantities like $\kappa^2 \tau_b$ and $\kappa^2 M^4$ to be smaller than order unity, we do not assume $\tau_f$, $\tau_o$ or $M^4$ to be particularly small relative to one another, and we wish to identify when the low-energy 4D curvature is set by scales that are much smaller than these.

What is important for what follows is that the choices \pref{observerbrane} and \pref{fluxbrane} ensure that the classical brane actions are both independent of the bulk field $\phi$. Our choices also ensure there is no direct coupling between the brane-localized fields $\psi$ and the bulk Maxwell field, $\cF_{\ssM\ssN}$.

\subsection{Bulk-brane interactions}

We next turn to the bulk configurations to which these two branes give rise. In what follows it suffices to focus on solutions that are maximally symmetric in the four on-brane directions, and are symmetric under rotations in the extra dimensions about the two brane positions. This leads to the following \emph{ans\"atze} for the metric and Maxwell field:
\be
\label{bulk-brane system: ansatz}
 \exd s^2 = \exd \rho^2 + e^{2B} \exd\theta^2 + e^{2W} \,
 \hat g_{\mu\nu} \exd x^\mu \exd x^\nu
 \quad \hbox{and} \quad
 \cA = \cA_\theta \, \exd \theta \,,
\ee
where $\hat g_{\mu\nu}(x)$ is a maximally symmetric metric, and the functions $W$, $B$, $\phi$ and $\cA_\theta$ depend only on $\rho$.

In this case the bulk field equations reduce to
\ba \label{eomansatz}
 \left( e^{-B + 4W} e^{-\phi} \cA_{\theta}' \right)' &=& 0
 \qquad(\cA_\theta) \label{Ade} \nn\\
 \left( e^{B+4W} \phi' \right)' - \left( \frac{2g_\ssR^2}{\kappa^2} \, e^\phi
 - \frac12\, \kappa^2 \cQ^2 \, e^\phi e^{-8W} \right) e^{B+4W} &=& 0 \qquad(\phi) \nn\\
 4 \Bigl[ W''+(W')^2 \Bigr] + B'' + (B')^2 + (\phi')^2 + \frac34 \, \kappa^2 \cQ^2
 \, e^\phi e^{-8W} + \frac{g_\ssR^2}{\kappa^2} \, e^\phi &=& 0 \qquad(\rho\rho) \\
 B'' + (B')^2 + 4W'B' + \frac34 \, \kappa^2 \cQ^2 \, e^\phi e^{-8W}
 + \frac{g_\ssR^2}{\kappa^2} \, e^\phi &=& 0 \qquad(\theta\theta) \nn\\
 \frac14 \, e^{-2W} \hat R + W'' + 4(W')^2 + W'B' - \frac14 \, \kappa^2\cQ^2 \, e^\phi e^{-8W}
 +\frac{g_\ssR^2}{\kappa^2} \, e^\phi &=& 0  \,,\qquad(\mu\nu) \nn
\ea
where primes denote differentiation with respect to the coordinate $\rho$. The first of these can be integrated once exactly, introducing an integration constant, $\cQ$:
\be \label{Afirstintegral}
 \cF_{\rho\theta}  = \cA_\theta'
  = \cQ \, e^\phi e^{B-4W}  \,.
\ee

Evaluated with this {\em ansatz}, the brane action, eq.~\pref{BraneFluxCoupling}, becomes
\be \label{BraneFluxCouplingAnz}
 S_\mathrm{branes} = - \sum_{b=o,f} \int d^4x \sqrt{- \hat g_4} \;
 e^{4W} L_b \,,
\ee
where $L_b$ is given for each brane in terms of $T_b$ and $\Phi_b$ by
\be
 L_b := T_b - \Phi_b \, e^{-B} \cF_{\rho\theta} + \cdots
 = T_b - \cQ \Phi_b \, e^{\phi}\, e^{-4W} + \cdots\,.
\ee

\subsubsection*{Brane matching conditions}

The brane-bulk couplings impose a set of boundary conditions on the derivatives of the bulk fields in the near-brane limits, that are the generalization to codimension-2 of the more familiar Israel junction conditions \cite{IJC} of codimension-1. The precise conditions were recently worked out for codimension-2 branes \cite{uvcaps, TNc2B, BBvN} (see also \cite{otheruvcaps, PST}), and state:
\be \label{matching}
 \Bigl[ e^{B} {\phi}' \Bigr]_{\rho_b}
 = \frac{\partial \cL_b}{\partial \phi} \,, \quad
 \Bigl[ e^{B} W' \Bigr]_{\rho_b} =  \cU_b
 \quad \hbox{and} \quad
 \Bigl[ e^B B'-1 \Bigr]_{\rho_b}
 = - \Bigl[ \cL_b +  3\,\cU_b \Bigr] \,,\nn
\ee
where both sides are evaluated in the near-brane limit,\footnote{An important complication for codimension-2 branes over codimension-1 is that both sides of eqs.~\pref{matching} generically diverge in the near-brane limit; requiring a renormalization of the brane action \cite{Bren}. (This renormalization turns out to be unnecessary in the special case of D7-branes in Type IIB supergravity \cite{BBvN}.)} $\rho \to \rho_b$, and as before primes denote differentiation with respect to $\rho$. The quantities $\cL_b$ and $\cU_b$ appearing here are
\be \label{TUdefs}
 \cL_b := \frac{\kappa^2 L_b}{2\pi}
 \quad \hbox{and} \quad
 \cU_b :=  \frac{\kappa^2}{4\pi} \left(
 \frac{\partial L_b}{\partial g_{\theta\theta}} \right) \,.
\ee
Notice that it is not necessary to know how $L_b$ depends on $g_{\theta \theta}$ in order to evaluate $\cU_b$, because the bulk field equations require $\cU_b$ must satisfy the constraint \cite{uvcaps, TNc2B, BBvN}
\be
\label{curvature constraint}
  4\cU_b \Bigl[ 2 - 2\cL_b - 3 \, \cU_b \Bigr]
  - (\cL_b')^2 \simeq 0 \,,
\ee
where $\cL_b' = \partial \cL_b / \phi$, and so
\begin{equation} \label{Ubsoln}
 \cU_b = \frac{1}{3} \left[ \left( 1
 - \cL_b\right) - \sqrt{ \left(1 - \cL_b\right)^2 -
 \frac{3}{4} \, (\cL_b')^2 } \right]
 \simeq \frac{ (\cL_b')^2 }{ 8 (1 - \cL_b)^2 } + \cdots \,,
\end{equation}
where the root is chosen so $\cU_b \to 0$ when $(\cL_b')^2 \to 0$.

The corresponding boundary condition for the Maxwell field implies that in a coordinate patch containing each source brane, eq.~\pref{Afirstintegral} integrates to (see also Appendix \ref{app:branefluxq}) \cite{susybranes}
\ba
 \cA_\theta(\rho) &=& \left( \frac{\Phi_o}{2\pi} \right) e^{\phi_o}
 + \cQ \int_{\rho_o}^{\rho}
 \exd \tilde\rho  \; e^{\phi+B-4W} \quad\;\; \hbox{observer brane} \nn\\
 &=& - \left( \frac{\Phi_f}{2\pi} \right) e^{\phi_f}
 + \cQ \int_{\rho_f}^{\rho} \exd \tilde\rho
 \; e^{\phi+B-4W}  \quad \hbox{flux brane} \,,
\ea
where $\Phi_o := \lim_{\rho \to \rho_o} \Phi_o[\phi(\rho)]$ --- appropriately renormalized \cite{Bren} --- and so on.

Requiring these two solutions to differ by a gauge transformation, $g^{-1} \partial_\theta \Omega$, on regions of overlap between the two patches implies the flux-quantization condition
\be \label{fluxquantzn}
 \frac{n}{g} = \frac1{2\pi} \left( {\Phi_o \, e^{\phi_o}
 + \Phi_f \, e^{\phi_f}} \right)
 + \cQ \int_{\rho_o}^{\rho_f} \exd \rho \; e^{\phi+B - 4W} \,.
\ee
This identifies $\Phi_{\rm tot} = \sum_b \Phi_b \, e^{\phi_b}$ as the part of the total magnetic flux carried by the branes \cite{SLED1}.

\subsection{Explicit solutions}

A great many explicit solutions to the above field equations and boundary conditions have been found, starting almost 30 years ago \cite{SSs}. Some of these are known at the linearized level \cite{stability,susybranes}, while others are exact solutions \cite{SLED1, GGP, WBW6D, GGPplus, 6DdS, TimeDep, Swirl, MultiBranes}. Many of these solutions provide explicit compactifications from six to four dimensions, and provide among the earliest examples of flux-stabilized compactifications.

For the present purposes, what is most interesting about the exact solutions is that the most general solutions are known \cite{SLED1, GGP, WBW6D, GGPplus} for the special case where the dilaton's radial derivative, $\phi'$, tends to zero at both brane positions. As is clear from the boundary conditions, eqs.~\pref{matching}, these solutions are the ones appropriate for the case where the brane actions do not depend on $\phi$: $\partial L_b/\partial \phi = 0$. What is remarkable about these solutions is that for {\em all} of them the on-brane geometries are flat: $\hat g_{\mu\nu} = \eta_{\mu\nu}$. We now briefly summarize these solutions in more detail.

\subsubsection*{Rugby balls}

A particularly simple situation is the special case where the dilaton is constant, $\phi = \varphi_0$, since then the solution is very easy to visualize: a rugby ball, sourced by two branes \cite{SLED1}:
\ba \label{symmansatz}
 \exd s^2 &=& e^{-\varphi_0} \left[ \exd \hat \rho^2
 + \alpha^2 \ell^2 \sin^2 \left( \frac{\hat\rho} \ell \right)
 \exd\theta^2 \right] + \hat g_{\mu\nu} \exd x^\mu \exd x^\nu \nn\\
 \cF_{\rho\theta} &=& \cF_{\hat\rho\theta} \, e^{-\varphi_0/2}
 = \cQ e^{\varphi_0/2} \alpha \ell \sin \left( \frac{\hat\rho} \ell \right) \,.
\ea
With this metric the volume of the extra dimensions is
\be \label{XDvolume}
 \cV_2 = 4 \pi \alpha \ell^2 e^{-\varphi_0} \,,
\ee
showing that the flux-stabilization fixes the extra-dimensional volume in terms of the scalar-field value, $\varphi_0$.

In these coordinates\footnote{Notice the coordinate rescaling $\rho := e^{-\varphi_0/2} \hat\rho$ between this solution and the {\em ansatz} of eq.~\pref{bulk-brane system: ansatz}.} the two source branes for this geometry are situated at $\hat\rho_o = 0$ and $\hat\rho_f = \pi \ell$. This geometry has a conical singularity at these points, characterized by the defect angle $\delta = 2\pi(1 - \alpha)$. In the special case $\alpha = 1$ the extra-dimensional geometry is a sphere \cite{SSs}. For these rugby-ball solutions the matching conditions, eqs.~\pref{matching}, degenerate to a relationship between the deficit angle and the lagrangian density, $L_o = L_f = L$, of the two source branes,
\be
 1 - \alpha = \frac{\kappa^2 L}{2\pi} \,,
\ee
as expected from other approaches \cite{TvsA}.

The equations of motion determine the values of $\cQ$ and $\ell$ as well as the curvature of the on-brane directions:
\be \label{rugbyrelns}
 \cQ = \pm \frac{2 g_\ssR}{\kappa^2} \qquad
 \ell = \frac{\kappa}{2 g_\ssR}  \quad \hbox{and} \quad
 \hat g_{\mu\nu} = \eta_{\mu\nu} \,,
\ee
while the flux quantization condition implies
\be \label{rugbyfluxquantwbranes}
  \frac{n}{g} = \frac{\alpha}{g_\ssR}
  + \left( \frac{\Phi_o + \Phi_f}{2\pi} \right) e^{\varphi_0}
 \,,
\ee
where the last equality uses eqs.~\pref{rugbyrelns}.

The interpretation of this last equation differs according to whether or not $\sum_b \Phi_b \, e^{\phi_b}$ depends on $\varphi_0$. If not --- such as in the scale-invariant case where $\Phi_b \propto e^{-\phi_b}$ ---  then eq.~\pref{rugbyfluxquantwbranes} must be regarded as a constraint on the parameters of the brane action which, if not satisfied, is an obstruction to the existence of solutions satisfying our assumed symmetry {\em ansatz}. But if $\sum_b \Phi_b \, e^{\phi_b}$ depends on $\varphi_0$ then this equation can be read as determining the value of $\varphi_0$, which is not fixed by any of the other field equations.

These properties have a simple interpretation from the point of view of the low-energy 4D effective theory \cite{susybranes, BvN}. The reason $\varphi_0$ is not fixed by the other equations is because it is the parameter that labels the one-parameter family of solutions whose existence is guaranteed by the scale invariance, eq.~\pref{classscaleinv}, of the classical bulk field equations. If both $T_b$ and $\Phi_b \, e^{\phi}$ do not depend on $\phi$ the brane couplings do not break this scale invariance, implying the classical low-energy potential for $\varphi_0$ must have an exponential form, $V_{\rm eff} = A e^{2\varphi_0}$. In this case there are two situations: ($i$) $\varphi_0$ labels a flat direction (and so is undetermined by the field equations) if $A = 0$, or ($ii$) $\varphi_0$ necessarily runs away to $\pm \infty$ if $A \ne 0$ (and so for finite $\varphi_0$ there are no solutions to the equations that are maximally symmetric in the on-brane directions). As is shown in \cite{susybranes} (see also Appendix \ref{app:4dpotential}), eq.~\pref{rugbyfluxquantwbranes} corresponds in the low-energy theory to the condition $A = 0$.

However, in the case of interest -- {\em c.f.} eqs.~\pref{observerbrane} and \pref{fluxbrane} -- $\sum_b \Phi_b \, e^{\phi}$ {\em does} depend on $\phi$, and so the brane-bulk couplings break the bulk scaling symmetry. In this case $\varphi_0$ appears only in eq.~\pref{rugbyfluxquantwbranes}, which should be read as fixing its value. From the point of view of the low-energy 4D theory (details in Appendix \ref{app:4dpotential}), $\varphi_0$ gets fixed because the breaking of scale invariance lifts the flat direction, through a Goldberger-Wise-like \cite{GW} stabilization mechanism for codimension-2 branes. In this case all of the field equations are not satisfied unless $\varphi_0$ is chosen to minimize this potential.

\subsubsection*{Unequal tensions}

Because the above rugby ball solutions have equal defect angles at both brane positions, they only describe situations where the two branes have precisely equal tensions. But the general solutions that apply when $\phi' \to 0$ at both branes are known \cite{WBW6D,GGP,GGPplus}, including those having two unequal brane tensions, which we now describe.

In this case the metric can be written
\be \label{sugraansatz}
 \exd s^2 = \cW^2 \, \hat g_{\mu\nu} \, \exd x^\mu \exd x^\nu +
 a^2 e^{-\varphi_0} \Bigl( \cW^8 \exd \eta^2 + \exd \theta^2 \Bigr) \,,
\ee
where $a = a(\eta)$, $\cW = \cW(\eta)$ and (as before) $\hat g_{\mu\nu}$ is a maximally symmetric on-brane geometry. The dilaton is similarly taken to depend only on $\eta$, $\phi = \phi(\eta)$, and the Maxwell field is given by $\cA_\theta = \cA_\theta(\eta)$, so that
\be
 \cF_{\eta\theta} = Q \, a^2 \, e^\phi  e^{- \varphi_0} \,,
\ee
with $Q$ an integration constant. With these choices proper distance along the direction between the branes is $\exd \rho = a \cW^4 \exd \eta \, e^{-\varphi_0/2}$, the proper circumference of a circle along which $\theta$ varies from zero to $2\pi$ at fixed $\eta$ is $\cC = 2 \pi a(\eta)\, e^{-\varphi_0/2}$, and the extra-dimensional volume is $\cV_2 = 2\pi e^{-\varphi_0} \int \exd \eta \, a^2 \cW^4$. In particular, when $\varphi_0$ is moderately large and negative --- so the bulk coupling satisfies $e^{\varphi_0} \ll 1$ --- then the `radius' defined by $\cV_2 = r^2$ can become exponentially large: $r^2 \propto e^{-\varphi_0}$.

The general solution to the bulk field equations having only conical defects is known explicitly for these variables, given by $\hat g_{\mu\nu} = \eta_{\mu\nu}$ together with
\ba \label{GGP}
    e^\phi &=& \cW^{-2} e^{\varphi_0} \nonumber\\
    \cW^4 &=& \left( \frac{\kappa^2 Q }{2g_\ssR} \right)
    \frac{\cosh[ \lambda(\eta - \eta_1)]}{\cosh[ \lambda
    (\eta - \eta_2)]} \\
    \hbox{and} \qquad
    a^{-4} &=& \left( \frac{2g_\ssR\kappa^2 Q^3}{\lambda^4}
    \right) \cosh^3[\lambda(\eta - \eta_1)]
    \cosh[ \lambda(\eta - \eta_2)]  \,,\nonumber
\ea
showing that $\eta_2 - \eta_1$, $\lambda$, $\varphi_0$ and $Q$ are the independent integration constants.

The position of the two source branes in these coordinates is $\eta \to \pm \infty$. Since the near-brane limit of the proper distance is
\be
 \exd \rho = \mp e^{-\varphi_0/2}a \cW^4 \, \exd \eta \to \mp C \, e^{\mp \lambda
 \eta} \, \exd \eta \,,
\ee
the defect angle in the geometry as $\eta \to \pm \infty$ turns out to be
\be
 \alpha_\pm := \left( \frac{2 \lambda g_\ssR}{\kappa^2 Q} \right)
 e^{\mp \lambda( \eta_2 - \eta_1) }\,.
\ee
The product of these last two expressions show how the integration constant $Q$ is fixed in terms of the tensions of the two branes:
\be
 \alpha_+ \alpha_- = \left( \frac{2 \lambda g_\ssR}{\kappa^2 Q} \right)^2 \,.
\ee
It is fixed in this way because it must be adjusted to ensure that the solution to the dilaton field equation is consistent with the boundary condition that $\phi' \to 0$ at both branes. Once this is done the solutions have three independent parameters that may be dialed: the two tensions (or defect angles) and the parameter $\varphi_0$ that labels the orbit of the classical scale symmetry.

The flux-quantization condition is found by computing $\cA_\theta(\eta)$ near the brane at $\eta \to \pm \infty$, giving
\ba
 \cA^\pm_\theta &=& \left( \frac{\Phi_o}{2\pi} \right) e^{\phi_o} +
 \frac{\lambda}{\kappa^2 Q} \Bigl\{
 \tanh[ \lambda (\eta - \eta_1) ] + 1 \Bigr\}
 \quad\;\; \hbox{observer brane} \nn\\
 &=& - \left( \frac{\Phi_f}{2\pi} \right) e^{\phi_f} +
 \frac{\lambda}{\kappa^2 Q} \Bigl\{
 \tanh[ \lambda (\eta - \eta_1) ] - 1 \Bigr\}
  \quad \hbox{flux brane} \,,
\ea
and so flux quantization becomes
\ba \label{warpedquantization}
 \frac{n}{g} &=& \frac{2 \lambda}{\kappa^2 Q} + \frac1{2 \pi}
 \left( {\Phi_o \, e^{\phi_o} + \Phi_f \, e^{\phi_f} } \right) \nn\\
 &=&  \frac{(\alpha_+ \alpha_-)^{1/2}}{g_\ssR}
 + \frac{1}{2 \pi} \left( \frac{\Phi_o}{\cW^2_o}
 + \frac{\Phi_f}{\cW^2_f} \right) e^{\varphi_0} \,,
\ea
where $\cW_0^4 = \lambda/\alpha_+$ and $\cW_f^4 = \lambda/\alpha_-$.
Notice this reduces to the rugby-ball quantization condition in the limit that\footnote{Notice that in the equal-tension limit the warp factor at the brane position is $\cW_o^4 = \cW_f^4 = \lambda/\alpha$, which was set to one in the rugby-ball solution by rescaling the coordinates $x^\mu$. } $\alpha_+ = \alpha_- = \alpha$.

\subsubsection*{Response to brane perturbations}

Crucial to what follows is what happens to these solutions when properties of the source branes are varied. Most importantly, the above solutions require their source branes to satisfy two separate conditions:
\ba \label{twoconditions}
 &&(i) \quad \frac{\partial L_b}{\partial \phi} = 0 \\
 &&(ii) \quad \hbox{flux quantization ({\em i.e.} eq.~\pref{warpedquantization})} \,.\nn
\ea
Notice in particular that it is {\em not} necessary to require $L_o = L_f$, which just corresponds to the special case of rugby-ball solutions.

These conditions provide the motivation for the choices made for the brane lagrangians given above --- eqs.~\pref{observerbrane} and \pref{fluxbrane}. The $\phi$-independence of both $T_b$ and $\Phi_b$ is designed so that both branes do not couple to $\phi$, ensuring $\partial L_o / \partial \phi = \partial L_f / \partial \phi = 0$ as required by condition $(i)$. But because these choices imply that the flux-quantization condition depends on $\varphi_0$, condition ($ii$) is automatically satisfied for an appropriate choice $\varphi_0 = \varphi_\star$. Using $\Phi_o = 0$ and $\Phi_f = \mu$:
\be
 e^{\varphi_\star} = \frac{2\pi \cW^2_f}{\mu}
 \;  \left[ \frac{n}{g} - \frac{(\alpha_+ \alpha_-)^{1/2}}{g_\ssR} \right] \,.
\ee

This adjustment of $\varphi_0$ also has an energetic interpretation. This can be shown explicitly for small perturbations about rugby-ball geometries (see  Appendix \ref{app:4dpotential} for details), for which condition ($ii$) can be seen to be the condition for minimizing the brane-generated scalar potential that lifts the flat direction for $\varphi_0$ in the low-energy 4D effective theory.

The same thing can also be shown beyond the linearized approximation. On general grounds, for the system studied here the effective 4D scalar potential responsible for the on-brane curvature is given by \cite{uvcaps, BBvN}
\be \label{Veffdef}
 V_{\rm eff}(\varphi_0) = V_{\rm brane}(\varphi_0) + V_{\rm bulk}(\varphi_0) \,,
\ee
where the `bulk' contribution is given by evaluating the bulk action at the bulk solution generated by the source branes,
\ba
 \sqrt{-g_4} \; V_{\rm bulk}
 &=& - \int \exd^2 x \cL_{\rm bulk} = - \frac{1}{2\kappa^2} \int \exd^2x \sqrt{-g} \; \Box \phi \nn\\
 &=& \frac{2\pi}{2 \kappa^2} \sum_b \sqrt{-g} \; n_\ssM \partial^\ssM \phi = \frac12 \, \sqrt{-g_4} \; L_b' \,.
\ea
where we use the general result, eq.~\pref{BulkSclass}, $n_\ssM$ is the normal vector directed {\em into} the bulk, evaluated at the position of each brane, and we use the dilaton matching condition, eq.~\pref{matching}, to trade $n_\ssM \partial^\ssM \phi$ for $L_b' = \exd L_b/\exd \phi$ evaluated at the brane.

The `brane' contribution to eq.~\pref{Veffdef} is similarly given by the sum of the brane action, $L_b$, and a `Gibbons-Hawking' contribution, for each brane. This leads to
\be
 V_{\rm brane} = \sum_b U_b \,,
\ee
with $U_b$ given by eqs.~\pref{TUdefs} and \pref{Ubsoln}. Notice in particular that near a zero of $L_b'$ the function $U_b$ vanishes quadratically:
\be
 \cU_b = \frac{\kappa^2 U_b}{2\pi}
 \simeq \frac{ (\cL_b')^2 }{ 8 (1 - \cL_b)^2 } + \cdots \,,
\ee
where $\cL_b = \kappa^2 L_b/2\pi$.

Combining terms, the total brane-generated effective potential becomes
\be \label{VeffvsU}
 V_{\rm eff} = \sum_b \( U_b + \frac12 \, \frac{\partial L_b}{\partial \phi} \) \,.
\ee
Notice in particular how $V_{\rm eff}$ vanishes whenever $L_b'$ does. This is required by consistency since all exact solutions are known with $\phi' \to 0$ at the branes (as is the case whenever $L_b' = 0$), and the on-brane geometry of all of these solutions is flat. Furthermore, since $U_b$ is quadratic in $\kappa^2 L_b'$ when $L_b'$ is small, it is only the second term in the sum in eq.~\pref{VeffvsU} that contributes in a linearized deviation away from these flat solutions, consistent with the explicit linearized analysis of Appendix \ref{app:4dpotential} and refs.~\cite{susybranes, BvN}. The $U_b$ term provides the generalization of the brane-generated potential beyond linear order, that is exact (up to classical level in the bulk physics) \cite{uvcaps, TNc2B, BBvN}.

\section{Loop effects}

In order to have a technically natural cosmological constant, it is not enough just to have a vanishing classical contribution. Since the cosmological constant problem is in essence a quantum problem, the problem hasn't become hard until loop effects are included. Generically, because the vacuum energy has low (mass) dimension, it is the largest mass scale that can appear in the loop that is the most dangerous. In the present instance there are two separate kinds of loop effects to distinguish: those involving only particles localized on the brane (which we imagine also includes all the known standard-model loops); and those that also involve virtual contributions from the bulk supergravity. We briefly discuss each in turn.

\subsection{Brane loops}

Consider first those loops involving only brane-bound states. For realistic brane-world models these include loops of all ordinary Standard Model particles. Neglecting (for the moment) bulk loops amounts to asking how the bulk and on-brane geometry classically respond to brane-loop-generated changes to the brane action.

Now comes the main point. What is important for these purposes is the observation that brane loops cannot in themselves invalidate the two conditions, \pref{twoconditions}, given that these are satisfied by the classical brane action ({\em i.e.} such as by eq.~\pref{BraneFluxCoupling} with eqs.~\pref{observerbrane} and \pref{fluxbrane}). That is, a sufficient condition for obtaining zero on-brane curvature (at the bulk classical level) is the {\em absence} of a coupling between the bulk dilaton, $\phi$, and the branes, since this ensures the validity of both conditions $(i)$ and $(ii)$ \cite{uvcaps, TNc2B}.

{}From this point of view the effects of brane loops can be regarded as generic $\cO(M^4)$ perturbations to the initial brane function $T_o$. For the model considered brane loops alone also cannot modify $\Phi_o$ because brane fields do not initially couple to the bulk gauge potential, $\cA_\ssM$. The assumed absence of heavy brane-localized fields on the flux brane, together with the physical separation between the observer and flux branes, similarly ensures that brane loops cannot modify\footnote{That is, the only influence at this order between the two branes is due to the classical response of the bulk fields, which are computed exactly in the above solutions, and do not correspond to changes to the flux-brane action.} flux-brane properties like $T_f$ or $\Phi_f$.

The upshot is that brane loops can only renormalize the brane actions (and in the model considered here, only for the observer brane) in a $\phi$-independent way. But this does not change the bulk response since we in any case did not assume anything special about the typical energy scale for $L_o$ when inferring the flatness of the on-brane geometry.

\subsection{Bulk loops}

Since brane loops cannot lift the flatness of the on-brane directions, the dominant corrections come from bulk loops. And these can come in a number of varieties, depending on whether or not the bulk states in the loop are short- or long-wavelength. The purpose of this section is to recap earlier arguments \cite{uvcaps, TNc2B} that the contribution of bulk loops to the low-energy scalar potential can be naturally of order $m_\KK^4$ in supersymmetric theories.

We first estimate the generic size of bulk loops in non-supersymmetric theories, and then how bulk supersymmetry changes these estimates.

\subsubsection*{Loops involving massless 6D fields}

On dimensional grounds the contributions of massless 6D fields to the low-energy 4D scalar potential is of order $\delta V_{\rm eff} \simeq m_\KK^4 \propto 1/r^4$, and various contributions of this type have been explicitly calculated for specific extra-dimensional geometries as Casimir energy calculations \cite{Casimir, CENontori, CETori, 6Dbulkloop}. Because the bulk states that dominate in the loop have wavelengths comparable to the size of the extra dimensions, this contribution to $V_{\rm eff}$ need not have a local interpretation from the point of view of the extra dimensions.

We now argue that order $m_\KK^4$ contributions are the generic size when the bulk is supersymmetric, since (unusually) the contribution of heavier fields is not larger than this.

\subsubsection*{Massive 6D states}

The Casimir energy contributed by 6D states of mass $m$ has also been computed \cite{CENontori, CETori} for simple extra-dimensional geometries. In general this depends in a complicated way on the dimensionless ratio $m/m_\KK$, but the simplifies considerably when $m \gg m_\KK$. The simplification arises because in this limit the wavelength that dominates the loop is much shorter than the size of the extra dimensions, leading to a result that can be described by a local contribution to the higher-dimensional effective action. This simplification allows a very general calculation \cite{UVsensitivity} of the contributions of heavy fields to the low-energy theory to be performed, using general tools \cite{GilkeyDeWitt} for studying the small-distance singularities in correlation functions on curved space.

There are two kinds of such local contributions that massive loops can generate. Quantum fluctuations that take place further than $\cO(m^{-1})$ from the branes are described by local contributions to the bulk action, integrated over the full 6D spacetime. Those that occur nearer to the branes themselves can also contribute local 4D corrections to the brane action. We consider each of these in turn.

\medskip\noindent{\em Far from the brane}

\medskip\noindent
The contributions in the bulk can be organized in a derivative expansion, leading to the schematic terms
\ba \label{localform}
 - \frac{ \delta \cL_{\rm eff}}{\sqrt{- g_6}} &=& a_1 m^6 + m^4 \Bigl[
  b_1 R + b_2 ( \partial_\ssM \phi \,
 \partial^\ssM \phi ) + \cdots \Bigr] + m^2 \Bigl[ c_1 R^2 + c_2
 ( \partial_\ssM \phi \, \partial^\ssM \phi)^2 + \cdots \Bigr] \nn\\
 && \qquad \qquad \qquad
 + \log \left( \frac{m^2}{\mu^2} \right) \Bigl[ d_1 R^3 + d_2
 R ( \partial_\ssM \phi \, \partial^\ssM \phi)^2 + \cdots \Bigr]+ \cdots  \,,
\ea
where all possible terms containing a fixed number of derivatives are included, for each of which the coefficients $a_i(\phi)$, $b_i(\phi)$, $c_i(\phi)$ and $d_i(\phi)$ are calculable (and generically nonzero) for any given choice for the heavy fields circulating in the loops \cite{UVsensitivity}.\footnote{For simple toroidal examples it can happen that the vanishing of ${R^\ssM}_{\ssN \ssP \ssQ}$ and $\partial_\ssM \phi$ in the background can imply that only the first of these survives, making the $\cO(m^6)$ contribution the only one that grows in the large-$m$ limit \cite{CETori}.} As indicated, in general these coefficients can be functions of the background scalar field, $\phi$.

The contribution of this kind of loop to the low-energy 4D potential for the zero-mode $\varphi_0$ may be estimated by replacing all derivatives by $1/r$, where $r$ is a measure of the extra-dimensional size, with the result integrated over the extra dimensions:
\be
 \delta V_{\rm eff} (\varphi_0) \simeq  \tilde a_1 m^6 r^2 + m^4 \Bigl[
  \tilde b_1 + \tilde b_2 + \cdots \Bigr] + \frac{m^2}{r^2} \Bigl[ \tilde c_1
  + \tilde c_2  + \cdots \Bigr] + \frac{1}{r^4} \,
  \log \left( \frac{m^2}{\mu^2} \right) \Bigl[ \tilde d_1  + \tilde d_2
  + \cdots \Bigr]+ \cdots \,,
\ee
where the coefficients $\tilde a_i$ through $\tilde d_i$ are proportional to $a_i$ through $d_i$, with numerical factors (and possibly logs of $m$ and $r$) arising from the details of the evaluation of the derivatives and the extra-dimensional integration. This shows that when $m \gg m_\KK \simeq 1/r$, it is the contributions involving $a_i$, $b_i$ and $c_i$ that are much greater than $\cO(m_\KK^4)$.

Here is where supersymmetry in the bulk plays its part. If we specialize to the classical level in the bulk, then there is no Casimir energy, $\delta \cL_{\rm eff} =0$, but the classical bulk lagrangian, $\cL_{\rm bulk}$ given by eq.~\pref{BulkAction}, itself has the form of eq.~\pref{localform}. And working to classical level in the bulk we know that in this case the work of the previous sections shows that the contributions to the low-energy potential cancel\footnote{Explicitly, for the rugby-ball solutions it is the coefficients of the $R$, $e^{-\phi} F^2$ and $e^\phi$ terms that cancel amongst themselves, which is possible because $e^\phi \propto 1/r^2$ for the classical solution.} to give $V_{\rm eff}^c = 0$.

Bulk loops change this, but their $\phi$-dependence is easy to establish because for supersymmetric theories the classical scale invariance implies $e^{2\phi}$ is the loop-counting parameter. In the frame where the classical lagrange density has the form $\cL_{\rm bulk} \propto \sqrt{- \check g} \; e^{-2\phi}$, the $\ell$-loop corrections obtained after integrating out heavy 6D states of mass $m$ are therefore proportional to $\delta \cL_\ell \propto \sqrt{- \check g} \; e^{2(\ell -1) \phi}$, which implies in Einstein frame $\cL_{\rm bulk} + \delta \cL_{\rm eff}$ is given by
\ba \label{localform2}
 - \frac{\cL_{\rm bulk} + \delta \cL_{\rm eff}}{\sqrt{- g_6}}
 &=& \left[ \frac{2 g_\ssR^2}{\kappa^4}  \, e^\phi + a_1 \, m^6 \, e^{3\phi}
  + \cO(e^{5\phi}) \right]\nn\\
  && \qquad + \left[ \frac{1}{2\kappa^2} + b_1 \, m^4 \, e^{2\phi}
   + \cO(e^{4 \phi})  \right] R + \cdots \nn\\
   && \qquad \qquad + \left[ c_1 \, e^{\phi} + \cO(e^{3\phi})  \right]
    m^2 \, R^2 + \cdots \nn\\
 && \qquad \qquad \qquad
 + \left[ d_1  + \cO(e^{2\phi}) \right] R^3  \log \left( \frac{m^2}{\mu^2} \right)
 + \cdots  \,.
\ea

Notice in particular that all of the corrections beyond the classical terms are at least $\cO(1/r^6)$ once evaluated with derivatives of order $1/r$ and using the classical flux-stabilization condition, $e^\phi \propto 1/r^2$. This ensures all such contributions to the 4D potential are at most of order $\delta V_{\rm eff} \propto 1/r^4 \simeq m_\KK^4$ for large $r$, as claimed. Evidence from explicit calculations for this supersymmetric suppression is also available for some kinds of compactifications \cite{UVsensitivity}.

Furthermore, validity of the semiclassical tools used here ensure the coefficient of proportionality of $1/r^4$ also cannot be large. For example, for the rugby ball if we define $r$ using the extra-dimensional volume, so $\cV_2 := r^2$, then eqs.~\pref{XDvolume} and \pref{rugbyrelns} imply
\be
 r^2 e^{\varphi_0} = 4 \pi \alpha \left( \frac{\kappa}{2 g_\ssR} \right)^2 \,,
\ee
and so, for example, $\cV_2 \, m^6  e^{3\varphi_0} = (\pi \alpha \, \kappa^2 m^2/g_\ssR^2)^3 (1/r)^4 \propto (m/M_g)^6 (1/r)^4$, where we take $g_\ssR^2 \simeq \kappa \simeq 1/M_g^2$ with $M_g$ the 6D gravity scale. The validity of the semiclassical approximation in the low-energy theory requires $m \ll M_g$, which keeps the coefficient of $1/r^4$ small.

More precisely, loops involving states with mass $\kappa^2 m^2 > 1$ would have to be computed in the UV completion of the low-energy supergravity. Although string theory provides a natural choice for this, we cannot yet compute these loops for the 6D supergravity studied here since its string-theoretic provenance is not yet known (see however \cite{CP, StrSLED}). We do know, however, that at such high energies there are a variety of mechanisms \cite{ubernat}, including supersymmetry and the general softening of UV dependence that string theory brings, that can suppress these contributions from the extreme ultraviolet.

\medskip\noindent{\em Near the brane}

\medskip\noindent
A similar discussion applies to quantum fluctuations of heavy bulk fields located near the branes. These also have a local interpretation if the bulk fields involved have masses $m \gg m_\KK$. Proximity to the brane allows such  loops to modify the brane lagrangian as well as the bulk one. Since each bulk loop comes with a factor of $e^{2\phi}$, near-brane loop effects are counted by also writing the brane lagrangians as a series in this variable,
\be
 T_b = T_b^{(0)} + T_b^{(1)} \, e^{2\phi} + \cdots \,,
 \quad \hbox{and} \quad
 \Phi_b = \Phi_b^{(0)} + \Phi_b^{(1)} \, e^{2\phi} + \cdots \,,
\ee
and so on. Such corrections are potentially dangerous because they clearly introduce a $\phi$-dependence to the brane action, and so violate condition $(i)$, above, that ensured the flatness of the on-brane directions.

The effect of integrating out very massive bulk states therefore corresponds to modifying both the brane and bulk actions as a series in $e^{2\phi}$. And the implications of the corrections to the brane action can then be estimated by following how these changes modify the bulk solutions, through the changes they induce in the bulk boundary conditions. This is evaluated in detail in Appendix \ref{app:4dpotential}, showing that the result is a contribution to the effective potential that is of order $e^{2\varphi_\star}$, where $\varphi_\star$ is the lowest-order value of the localized dilaton. The rest of the story is by now familiar: because $e^{2\varphi_\star} \propto 1/r^4$, the resulting contribution to the 4D potential is again $\delta V_{\rm eff} \simeq 1/r^4 \simeq m_\KK^4$, as claimed.

The upshot is this: brane loops in themselves cannot cause on-brane curvature because they cannot introduce a $\phi$-dependence of the brane action if this was absent at lowest order. Bulk loops can cause on-brane curvature, but the result corresponds to an effective 4D potential that is of order $\delta V_{\rm eff} \simeq m^4 e^{2\phi} \simeq 1/r^4$, and so is very small for large $r$ because the bulk coupling is so very weak in this limit. Although the calculation of contributions from loops arising from above the gravity scale remain beyond our present calculation reach, similar kinds of volume suppression are known to arise in other explicit large-volume string compactifications \cite{StrSLED, ubernat}.

\section{Conclusions}

We provide here an explicit model of brane-localized matter for which both brane-backreaction and fluxes play a role in stabilizing the size of the extra dimensions. Remarkably, the stabilization mechanism produces an on-brane curvature that is parametrically suppressed relative to the generic scales of masses that define the brane-localized tensions (including loops).

The model of interest involves a generic field theory (a proxy for the Standard Model, say) localized on a nonsupersymmetric codimension-two brane within a six-dimensional spacetime whose bulk dynamics is supersymmetric. The on-brane curvature is found to be of order $R \sim V/M_p^2$, where $V \sim m_\KK^4$. This is true even if the Kaluza-Klein scale, $m_\KK$, is much smaller than the generic particle mass, $M$, on the brane.

The small size of the low-energy effective potential is a consequence of a cancelation between the direct contributions of the brane and the contributions of the bulk to which the branes give rise. What is new in this paper is an explicit calculation of how the system responds to arbitrary small perturbations in brane properties, which confirms in detail the more general arguments \cite{WBW6D, uvcaps, TNc2B, BBvN, GGPplus, UVsensitivity, ubernat} that have emerged over the years from the SLED proposal \cite{SLED1, SLEDrevs}.

We believe our example provides a useful explicit particular realization of the general SLED proposal, but expect the result to apply more generally than just for the specific 6D supergravity considered here. The ingredients we believe to be necessary to suppress the brane curvature are:
\begin{itemize}
\item Codimension-two branes, for which the back-reaction on the surrounding bulk only varies logarithmically with distance, and so cannot be neglected even when comparatively far from the brane;
\item A higher-dimensional bulk, described by a supergravity that enjoys a classical scale invariance under which the metric scales by a constant factor while a bulk dilaton shifts. Such a scale invariance appears to be generic for many supergravities in six and higher dimensions. The bulk itself need not be required to be invariant under any of the supersymmetries.
\item A bulk-stabilizing flux that can be localized on at least one of the branes present.
\end{itemize}
In particular, we expect the mechanism to generalize to 3-branes localized within a bulk described by more generic 6D supergravities than the particular Nishino-Sezgin gauged supergravity studied here. But supersymmetry is crucial, since this is what allows the suppression of bulk loop effects.

We believe the models described here provide a context for understanding why the observed Dark Energy density is so much smaller than all of the other scales we know about in particle physics. Because the stabilization mechanism for the extra dimensions both explains the dark energy density and the electroweak hierarchy, these problems are related in this framework.

\subsection{Some observational implications}

Realistic applications to an effectively 6D world with large supersymmetric dimensions require $m_\KK \simeq 10^{-2}$ eV, which corresponds to $\cV_2/\kappa \simeq (M_g r)^{2} \simeq e^{-\varphi_\star} \simeq 10^{30}$. Notice flux-quantization gives the value of the stabilized dilaton by $e^{-\varphi_\star} \simeq g_\ssR \mu / \kappa^2 T \simeq 10^{30}$, where $T$ is a generic brane tension and $\Phi = \mu$ is the flux parameter on the flux-brane. The extra dimensions would be expected to be of order $m_\KK \simeq 10^{-2}$ eV, or $r \simeq 10$ microns, and the 6D gravity scale could be as low as\footnote{A gravity scale lower than this produces too much energy loss from supernovae \cite{ADD, MSLED}. Much stronger, but more model-dependent, bounds are also possible if extra-dimensional states can decay visibly (such as into photons) \cite{HR, SNothers}, but these bounds can be avoided if visible channels are swamped by invisible higher-dimensional ones \cite{ADD, MSLED} (some potential examples of which are discussed in \cite{StrSLED}).} $M_g \gsim 10$ TeV.

If this is how nature works we will soon know, since the SLED picture necessarily has many striking observational consequences, some of which are shared by the non-supersymmetric proposal for sub-eV dimensions \cite{ADD}. Since some of these are described in more detail elsewhere, we only briefly list some of the main ones here.
\begin{itemize}
\item {\em Deviations from Newton's Law:} Since the size of the Dark Energy density is set by the KK scale, the extra dimensions must generically be of order micron scales. Deviations from Newton's inverse-square law must arise once distances of order this size are probed. This is the smoking gun for the SLED scenario, since it cannot be avoided. Since only two dimensions can possibly be this large, the predicted change is a crossover to an inverse fourth power, although the precise shape depends somewhat on the details of the extra-dimensional shape \cite{SLEDforce}. Present bounds probe down to about 45 microns \cite{GRTests, InvSqTests} and so are getting close.
\item {\em String and gravity physics at the LHC:} Given the size of the extra dimensions, the measured strength of gravity dictates the gravity scale in the extra dimensions. The 6D gravity scale to which this points is of order tens of TeV (though astrophysics requires it to be no smaller than 10 TeV). The string scale and the KK scale for any other extra dimensions is then generically found to be lower than this \cite{MSLED, StrSLED}. This means that quantum gravity is becoming strong at LHC energies. This implies a variety of signals for the LHC, including excited string states for all Standard Model particles \cite{stringLHC}, new neutral gauge bosons \cite{newbosons}, energy loss into gravitons \cite{ADDpheno} and other particles \cite{MSLED, SUSYADD, Hugo} in the extra dimensions, and possibly black holes \cite{LHCBH} or other aspects of high-energy gravity \cite{HEGrav}. Even though supersymmetry is broken only at very low scales in the bulk, supersymmetry must be nonlinearly realized on any brane and so superpartners for ordinary particles (and so also the MSSM) are {\em not} predicted \cite{MSLED}. Results for new experimental searches at the LHC are even now starting to come out \cite{ColliderBoundsLHC, ColliderBoundsLHCth}.
\item {\em Dark Energy quintessence cosmology:} The same physics that makes the value of the potential, $\rho = V_\star$, small at its minimum (and thereby gives a small Dark Energy density) also makes the mass of the would-be zero mode very light: $m^2 \simeq \sqrt{V_\star}/M_p$ (and in an equally technically natural way). Since this is of order the present-day Hubble scale, Dark Energy phenomenology is that of a quintessence model rather than of a cosmological constant \cite{SLEDCosmo}. The same requirement that makes the on-brane curvature small --- the absence of a direct brane-dilaton coupling --- also ensures that the light scalar field naturally has quasi-Brans-Dicke couplings to brane matter. This means they can naturally evade tests of the equivalence principle \cite{GRTests}, but the couplings need not be small and so are potentially constrained by a variety of long-distance tests of General Relativity that bound scalar-tensor models \cite{STBounds, STBoundsUS}, as well as laboratory bounds on light scalars with an effective 2-photon coupling \cite{LabBounds}. Present-day bounds on deviation from GR in the solar system provide nontrivial constraints, but need not be fatal \cite{SLEDCosmo}. One reason for this is because the Brans-Dicke couplings of the light scalar turn out to be field dependent, and so can evolve cosmologically. For parts of parameter space \cite{SLEDCosmo} these couplings can be acceptably small in the solar system during the present cosmological epoch.
\item {\em Exotic sterile neutrino physics:} Although not absolutely required, the SLED scenario predicts there to be a variety of massless fermions in the extra dimensions, whose mass is protected to be small because they are related by supersymmetry to the graviton or bulk gauge fields. These fermions can mix with Standard Model neutrinos, leading to a potentially rich spectrum of sterile neutrino mixing \cite{SLEDnus} whose masses are naturally in the sub-eV range due to the large size of the extra dimensions \cite{LEDnus1, LEDnus2}.
\end{itemize}
There are likely even more consequences, since only the surface of what might be seen has yet been scratched. Should all of these be seen together, there could be little doubt about what is going on.

\subsection{Outstanding issues}

We now summarize potential challenges that these models remain to face.

First, it is an unpleasant --- though technically natural --- feature of the model that a large number must be inserted for $\mu$ as a parameter in the lagrangian in order to obtain a sufficiently low KK scale. This does not cause a problem with the approximations made, however, since it is only the combination $g_\ssR \mu e^{\varphi_\star} \simeq \kappa^2 T \lsim 0.1$ that appears in the brane lagrangian. $\varphi_\star$ also appears on its own in the bulk lagrangian, but the loop approximation in the bulk is under good control because the loop counting parameter there is $e^{2\varphi_\star} \simeq 10^{-60}$. We expect this feature is likely something that can be improved in more complicated examples, preferably with more explicit contact with a UV string construction, since most of the known 10-dimensional string compactifications having very large volumes \cite{LVS} generically obtain equally large volumes as are required here without having to dial in such small parameters. They do so because they predict the volume to arise as the exponential of another, much smaller, modulus, for which parameters of order 10 need be used.

Second, much could be gained if this picture could be properly embedded into a controlled UV completion, such as if it were obtained from an explicit string vacuum. Until this is done the contributions to $\rho$ from states in the far UV cannot be properly computed.

Third, SLED models face a variety of phenomenological challenges as well as opportunities. In particular, as mentioned above, strong bounds on light gravitationally coupled fields must be evaded in order to not conflict with known physics in the solar system. The cosmology of the universe before nucleosynthesis is also challenging, due to constraints from energy loss into the extra dimensions (together with stronger, but more model-dependent bounds that arise if extra dimensional fields can decay too frequently to visible states). The nature of inflationary cosmology is also unknown (see however \cite{BvNCosmo} for first steps towards an inflationary theory where extra dimensions evolve during inflation, allowing the gravity scale to be much higher during inflation than it is at present). We regard these to be model building challenges, but much easier to solve than is the cosmological constant problem itself.

\subsection*{Acknowledgements}

We wish to thank Markus Luty, Fernando Quevedo and Raman Sundrum for useful discussions about backreaction in codimension-two models. CB acknowledges the Kavli Institute for Theoretical Physics in Santa Barbara and the Abdus Salam International Center for Theoretical Physics for providing the very pleasant environs in which some of this work was performed. LvN thanks the Instituut-Lorentz for Theoretical Physics at Leiden University for their hospitality. Our research is supported in part by funds from the Natural Sciences and Engineering Research Council (NSERC) of Canada. Research at the Perimeter Institute is supported in part by the Government of Canada through Industry Canada, and by the Province of Ontario through the Ministry of Research and Information (MRI).

\appendix

\section{Localized brane fluxes}
\label{app:branefluxq}

An important role is played by brane-localized flux in the discussion of the main text, and in particular the choice of large values for $\Phi_b$. In this appendix we use a simple but explicit model of microscopic brane dynamics to explore how reasonable these choices might be. In particular, one might worry that microscopic details (like flux quantization) of brane-localized flux could obstruct its role in the relaxation mechanism for the low-energy curvature.

Ideally this question should be addressed within string theory, which provides the most likely UV completion. However we are handicapped by the lack of a controlled derivation of 6D gauged chiral supergravity from an explicit string vacuum (see however \cite{CP, StrSLED}). Instead, as a first step we model the codimension-2 brane with localized flux as a very small cylindrical codimension-1 brane situated at $\rho = \epsilon$ which surrounds the position of the codimension-2 brane at $\rho = 0$, along the lines of refs.~\cite{uvcaps, PST}. We regard this brane, together with a suitably smooth interior configuration for $\rho < \epsilon$, as a specific UV completion of the codimension-2 brane. Although this is unlikely to be a realistic microscopic realization of brane flux localization, it has the advantage of allowing an explicit examination of many of the consistency issues involved.

Consider therefore the following codimension-1 brane action,
\be
 S_5 = - \int_{\rho=\epsilon} \exd^5x \sqrt{-g_5} \;
 \Bigl[ Z_1(\phi) D_m \sigma D^m\sigma + T_1(\phi) \Bigr] \,,
\ee
describing a small cylinder at radius $\rho = \epsilon$, where $\sigma$ is a brane-localized St\"uckelberg field whose covariant derivative is
\be
 D_m\sigma = \pd_m \sigma + g_b \cA_m \,.
\ee
This is invariant under the gauge transformations
\be
 \cA_m \to \cA_m -\frac1g \, \pd_m \Omega
 \quad \hbox{and} \quad
 \sigma\to\sigma+\frac{g_b}{g} \, \Omega\,.
\ee
Here $g$ denotes the bulk gauge coupling while $g_b$ denotes a corresponding brane gauge coupling.

The presence of a field like $\sigma$ is important for stabilizing the size of the codimension-1 brane at a small but nonzero radius \cite{uvcaps, TNc2B, PST}. For $\epsilon$ sufficiently small the codimension-1 brane becomes effectively a codimension-2 brane, whose action can be found by dimensional reduction. Having a finite codimension-2 brace action in this limit generally requires the quantities
\be
 t_1(\phi) := \epsilon \, T_1(\phi) \quad \hbox{and} \quad
 z_1(\phi) := \epsilon \, Z_1(\phi) \,,
\ee
remain finite in the limit of small $\epsilon$.

In the region exterior to the brane, $\rho \ge \epsilon$, and in the presence of any bulk matter fields having charge $g$, the single-valuedness of the gauge group element, $e^{i \Omega}$, requires $\Omega(\theta + 2\pi) - \Omega(\theta) = 2\pi s$ for some integer $s$. The St\"uckelberg field can also wind nontrivially as a function of $\theta$ if its target space should be a circle,
\be
 \sigma(\theta + 2\pi) - \sigma(\theta) = 2 \pi n f \,,
\ee
for some nonzero integer $n$, where $2\pi f$ denotes the circumference of the target-space circle. This boundary condition is consistent with gauge transformations provided
\ba
 \sigma_\Omega(\theta + 2\pi) - \sigma_\Omega(\theta)
 &=& \sigma(\theta + 2\pi) + \frac{g_b}{g} \, \Omega(\theta + 2\pi)
 -\sigma(\theta) - \frac{g_b}{g} \, \Omega(\theta) \nn\\
 &=& 2\pi n f  + 2\pi s \;\frac{g_b}{g} \,,
\ea
is also an integer multiple of $2\pi f$. This is automatically true if the brane gauge coupling is quantized in units of the bulk gauge coupling: $g_b = k f g$ for some integer $k$. In this case because $\sigma_\Omega(\theta + 2\pi) - \sigma_\Omega(\theta) = 2 \pi (n + sk)f$ differs from $\sigma(\theta + 2\pi) - \sigma(\theta) = 2 \pi n f$, large gauge transformations (those with $s \ne 0$) map different choices for $\sigma$ boundary conditions into one another.

\subsubsection*{Brane equation of motion}

The equation of motion on the brane is
\be
 \pd_m \Bigl[ \sqrt{-g_5} \; Z_1(\phi) D^m \sigma \Bigr] = 0 \,,
\ee
and we are interested in solutions that depend on $\theta$ only. Since none of the bulk fields that appear in this equation of motion depend on $\theta$ this simplifies to
\be
 \pd_\theta D_\theta\sigma = 0 \,,
\ee
which has as solution
\be \label{sigma-eom}
 D_\theta \sigma= \pd_\theta\sigma+k f g \, \cA_\theta = C\,,
\ee
where $C$ is independent of $\theta$. However, since all of the bulk fields depend only on the radial coordinate $\rho$, they do not depend on any of the five on-brane directions (including $\theta$), and so $C$ can depend on any of them. In particular, $C$ can be a function of the dilaton, $\phi$.

The requirement that $\sigma$ be single-valued up to integer multiples of $2\pi f$ then means that
\be
 2\pi f n = \sigma(\theta + 2\pi) - \sigma(\theta) = \oint \exd \theta \; \pd_{\theta} \sigma = 2\pi  C
 - k f g \oint_{\rho = \epsilon}  \cA_{\theta}\, \exd \theta  \,,
\ee
which implies that $C$ is given in terms of the flux,
\be
 \Phi := \oint_{\rho = \epsilon} \cA_\theta \, \exd \theta  \,,
\ee
by
\be \label{Csoln}
 C = \left( n + \frac{kg}{2\pi} \, \Phi \right) f \,.
\ee
Notice that the transformation of $\Phi$ under large gauge transformations ({\em i.e.} those with $s \ne 0$) ensures that $C$ is invariant even though $n \to n + ks$.

We now specify in more detail the system interior to the codimension-1 brane, with the goal of deriving a second relationship between $C$ and $\Phi$, from which we may eliminate $C$. At first sight one might worry that any expression for $\Phi$ won't be gauge invariant, since $\Phi$ transforms under large gauge transformations. However once we match through to a smooth inner configuration the noncontractible loop whose topology underlies the existence of large gauge transformations disappears. The nontrivial large gauge transformations necessarily become singular somewhere once they are extended into the interior region.

\subsubsection*{The cylinder's interior}

We next specify the interior of the cylindrical brane, which we require to be everywhere smooth. We keep the action in the interior the same as in the bulk, apart from only one change: we take the dilaton potential to be
\be
 V=V_0 \, e^\phi\,,
\ee
for a general constant $V_0$. If we write $V_0 = 2g_\ssR^2/\kappa^4$, we effectively choose the value of $g_\ssR = g_\ssR^{\rm in}$ interior to the cylinder to differ from its value on the outside.

We take the interior solution to be the Salam-Sezgin solution,
\be
 \exd s^2 = e^{2W} \hat g_{\mu\nu} \, \exd x^\mu \exd x^\nu +
 e^{- \varphi_{\rm in}} \Bigl[ \exd \hat \rho^2 + e^{2B} \, \exd\theta^2 \Bigr] \,,
\ee
with
\be
 \phi = \varphi_{\rm in} \,, \qquad
 W = W_{\rm in} \quad \hbox{and} \quad
 \cF_{\rho\theta} = \cQ_{\rm in} e^{\varphi_{\rm in}/2} e^{B - 4W} \,,
\ee
where $\varphi_{\rm in}$ and $W_{\rm in}$ are constants, and
\be
 e^B = \ell_{\rm in}  \sin \left( \frac{\hat\rho - \hat\rho_c}{\ell_{\rm in}}
  \right)
 \,.
\ee
The center of the interior geometry is located at $\hat \rho = \hat \rho_c$, which need not be $\hat\rho=0$ due to our choice that the codimension-1 brane is located at $\rho  = \epsilon$ for both the exterior and interior geometries.

Like in the exterior geometry the equations of motion still imply
\be
 \ell_{\rm in} = \frac\kappa{2g_\ssR^{\rm in}} \,, \qquad
 \cQ_{\rm in} = \pm\sqrt{2V_0} = \pm \frac{2 g_\ssR^{\rm in}}{\kappa^2}
 \qquad {\rm and} \qquad \hat g_{\mu\nu}=\eta_{\mu\nu}\,,
\ee
which shows that we can dial the value of $V_0$ to achieve any desired flux for the interior gauge field. Choosing a gauge with $\cA_\theta(\rho_c) = 0$, with all other components of $\cA_\ssM$ vanishing, the gauge fields become
\ba
 \cF_{\rho\theta} &=& \cQ_{\rm in}  e^{ -4W_{\rm in} + \varphi_{\rm in}/2 }
 \left[ \ell_{\rm in} \sin \left(
 \frac{\hat\rho - \hat\rho_c}{\ell_{\rm in}}  \right) \right]
 \nn\\ \hbox{and so} \quad
 \cA_\theta(\rho) &=& \cQ_{\rm in} e^{-4W_{\rm in} }
 \ell_{\rm in}^2  \left[ 1 - \cos \left(
 \frac{\hat\rho - \hat\rho_c}{\ell_{\rm in}} \right) \right]
 \,. \\
\ea

\subsubsection*{Matching conditions}

Continuity of  the metric and dilaton at the brane location, $\rho = e^{-\varphi_{\rm in}/2} \hat\rho = \epsilon$, implies
\ba
 \varphi_{\rm in} = \phi(\epsilon) &=& \phi_b \nn\\
  W_{\rm in} = W(\epsilon) &=& W_b\nn\\
 \hbox{and } \quad
 e^{-\varphi_{\rm in}/2} \ell_{\rm in} \sin \left( \frac{\hat \epsilon - \hat \rho_c}{\ell_{\rm in}} \right)
 = e^{B_b} &=& \alpha_b \epsilon \,,
\ea
where $\hat \epsilon = \epsilon \, e^{\varphi_{\rm in}/2}$ and $\phi_b$, $B_b$ and $W_b$ are the (regulated) values of the dilaton and warping at the brane in the exterior bulk solution. From this we find the value of the gauge field at the brane is,
\be
\label{vectorpotential-boundary}
 \cA_\theta(\epsilon) \simeq
  \cQ_{\rm in} e^{-4W_{\rm in} }
 \ell_{\rm in}^2  \left[ 1 - \sqrt{ 1 - \left( \frac{ \alpha_b
 \epsilon\, e^{\phi_b/2}}{\ell_{\rm in}}  \right)^2 } \right] \,.
\ee

Next we impose the jump discontinuity of the gauge field across the brane position. For the above interior and exterior solutions and brane action, this reads
\be
 \cQ_{\rm in} - \cQ_{\rm out} = - \frac{e^{B+4W}}{\sqrt{- g_5}}
 \, \frac{\delta S_5}{\delta \cA_\theta}
  \simeq \frac{ 2 g_b C Z_1 e^{4W_b}}{\alpha_b \epsilon} \,.
\ee
Using this to trade $\cQ_{\rm in}$ for $\cQ_{\rm out}$ in the gauge potential, eq.~\pref{vectorpotential-boundary}, we see that when $\alpha_b \epsilon\, e^{\phi_b/2} \ll \ell_{\rm in}$ the result for $\cA_\theta(\epsilon)$ is proportional to
\be
 (\alpha_b \epsilon)^2 \, \cQ_{\rm in} = (\alpha_b \epsilon)^2
 \, \cQ_{\rm out} + \alpha_b\epsilon \,
 \Bigl( 2 g_b C Z_1 e^{4W_b} \Bigr) \,.
\ee
Although the first term on the right-hand side vanishes in the codimension-2 limit where $\epsilon \to 0$, the second term need not because the finiteness of the dimensionally reduced codimension-2 action obtained from $S_5$ requires $z_1  = \lim_{\epsilon \to 0} \epsilon \, Z_1$ be finite in this limit. In terms of this the brane-localized flux becomes
\be
 \Phi = \oint_{\rho = \epsilon} \cA_{\theta} \, \exd \theta
 = 2 \pi \cA_\theta(\epsilon)
 = 2 \pi \alpha_b g_b C z_1 \, e^{\phi_b} \,.
\ee

Combining this last result with eq.~\pref{Csoln} allows us to solve for $C$, giving the quantization condition
\be
 \frac C f \( 1 - \alpha_b g_b^2 z_1 \, e^{\phi_b} \) = n \,.
\ee
Equivalently, using this to eliminate $C$ from the flux gives
\be
 \frac{\Phi}{2\pi} = \frac{ n f  \alpha_b g_b  z_1\, e^{\phi_b} }{1
 - \alpha_b g_b^2 z_1 \, e^{\phi_b}} \,.
\ee
Notice that although this expression is quantized in the sense that it is proportional to an integer, it is also $\phi_b$-dependent through the quantity $z_1(\phi_b) \, e^{\phi_b}$. Furthermore, the regime of weak coupling and small derivatives has $g_b^2 z_1 \, e^{\phi_b} \ll 1$ and so we may approximate the denominator by unity, leading to a contribution to $\Phi$ that is proportional to $z_1 \, e^{\phi_b}$. (Intriguingly, if $g_b^2 z_1 \, e^{\phi_b}$ were instead large we would find the $\phi_b$-independent result $\Phi \to - 2\pi n f/g_b$, and so $g \Phi/2\pi \to - n/k$ would be quantized at rational values.)

In the special case that the brane does not break the bulk classical scale invariance then $T_1 \propto e^{\phi_b/2}$ and $Z_1 \propto e^{-\phi_b/2}$, so writing $\epsilon = \hat \epsilon \, e^{-\phi_b/2}$ we see that $t_1 = \epsilon \, T_1$ is $\phi_b$-independent and $z_1 = \epsilon \, Z_1 \propto e^{-\phi_b}$, as expected. This means $\Phi$ is independent of $\phi_b$, as is also argued to be true for the scale invariant case in the main text.

On the other hand, the case of most interest in the main text is where $\Phi = \mu \, e^{\phi_b}$, which corresponds to choosing $z_1$ to be $\phi_b$-independent. The above calculation then gives the coefficient, $\mu$, as
\be
 \mu = 2 \pi n \,\alpha_b  g_b z_1 f = 2 \pi nk \, \alpha_b g z_1 f^2\,.
\ee
In particular, we seek situations where $g_\ssR \mu$ is very large, while keeping $g_\ssR \mu \, e^{\phi_b}$ small. This we can arrange in several ways: ($i$) by making $f$ very large (so $\sigma$ takes values on a very large circle); ($ii$) by making the integers $k$ and/or $n$ very large; or ($iii$) by making $g_\ssR g z_1$ large. All of these choices come down to including a lot of current on the codimension-1 brane, as one might expect. Large $n$ means a very high gradient in $\sigma$, which can be interpreted as a lot of particles in the current. Large $k$ means a comparatively large coupling, $g_b$, which gives $\sigma$ a large charge. Finally, large $f$ gives both a large brane charge and a large gradient.

The main worry with these choices would be if they would indicate a failure of the low-energy derivative expansion, on whose validity the entire calculation rests. However since $\mu$ appears systematically in the brane action only through the combination $\mu \, e^{\phi_b}$ this expansion appears to be under control provided only that this product be small. $e^{\phi_b}$ also appears without factors of $\mu$ in the bulk action, but extremely small values of $e^{\phi_b}$ are there under control because this is the small quantity that controls the bulk loop expansion. In particular, there seems to be no consistency restriction on how large the parameter $f$ can be.

\section{The view from 4 dimensions}
\label{app:4dpotential}

In this section, we ask what the scalar potential is that lifts the flat direction parameterized by $\varphi_0$, as would be seen from the perspective of a brane-localized 4D observer. To do so we draw heavily on the results of \cite{susybranes}, which computes this potential for geometries that are perturbatively close to the rugby-ball geometries.\footnote{Beware the notational change, as ref.~\cite{susybranes} denotes $L_b$ by $T_b$; denotes $T_b$ by $\tau_b$; and denotes $\Phi_b$ by $\Phi_b e^{-\phi}$.}

Writing the brane action as
\be
 S_b = - \int \exd^4x \sqrt{-g_4} \; L_b = - \int \exd^4x \sqrt{-g_4} \( T_b + \frac12 \,\Phi_b \epsilon^{mn}F_{mn} \) \,,
\ee
we calculate the low-energy 4D effective potential in the special case that the tensions satisfy $T_b = \overline T + \delta T_b$, where $\delta T_b$ is much smaller than the (positive) average tension, $\overline T$. We further assume the background rugby-ball geometry satisfies
\be
 \frac{ng_\ssR}g = 1 - \frac{\kappa^2  \overline T}{2\pi} \,,
\ee
so that no background brane-localized flux is present.

We compute the response of the bulk to deviations $\delta T_b = T_b - \overline T$ by linearizing the bulk equations in $\delta T_b$ and $\delta \Phi_b$, obtaining the general solutions as a function of the parameter $\varphi_0$ that labels the orbits of the bulk scaling symmetry. The brane-bulk matching conditions define the boundary conditions that are then used to eliminate the integration constants in terms of brane properties. This allows the calculation of the stabilized value $\varphi_0 = \varphi_\star$ and the energy cost for deviations of $\varphi_0$ from $\varphi_\star$. In this way the features of the low-energy 4D potential can be mapped out \cite{susybranes}. (We emphasize that this linearization is not required for the arguments of the main text, for which the general exact classical solutions are known. This limit is simply one for which we can explicitly calculate the view in the 4D effective theory, to check our general results.)

Writing
\be
 \delta L_b = \Bigl( T_b -\overline T \Bigr) - \cQ \Phi_b \, e^{\varphi_0}
 = \delta T_b  - \cQ \Phi_b \, e^{\varphi_0} \,,
\ee
(which uses the result that $W = 0$ for the unperturbed rugby-ball geometry), the
stationary point, $\varphi_\star$, for the scalar zero mode turns out to be given by \cite{susybranes}
\be
\label{stabilization}
 0 = \sum_b \( \delta L_b + \frac12 \, \delta L_b'
 - \cQ\Phi_b \, e^{\varphi_\star} \) \,,
\ee
where prime denotes differentiation with respect to $\phi$. When $\delta T_b$ and $\Phi_b$ are both independent of $\varphi_0$ for all of the branes, then this simplifies to\footnote{Recall $\delta L_b'$ denotes differentiation with respect to $\phi$ with $\cA_\ssM$ and $g_{\ssM \ssN}$ fixed, and because $\cA_\ssM$ and $g_{\ssM \ssN}$ depend on $\varphi_0$ this is {\em not} the same as differentiating $\delta L_b$ with respect to $\varphi_0$.}
\be
\label{stabilization2}
 0 = \sum_b \Bigl( \delta T_b  - 2 \cQ \Phi_b \, e^{\varphi_\star} \Bigr) \,,
\ee
with solution
\be \label{appzeroethorder}
 e^{\varphi_\star} = \frac{\sum_b \delta T_b}{\sum_b 2 \cQ \Phi_b}   \,.
\ee

The Jordan frame potential of the low-energy effective 4D theory is shown in ref.~\cite{susybranes} to satisfy
\be \label{JFVDE}
  \( e^\varphi V_{\JF} \)'
 = \frac12 \, e^{\varphi}  \sum_b \( \delta L_b + \frac32 \, \delta L_b'
 - \cQ\Phi_b \)  \,,
\ee
which for $\phi$-independent $\delta T_b$ and $\Phi_b$ can be integrated to give
\be
 V_{\JF}(\varphi) = Ce^{-\varphi} + \frac12 \sum_b \Bigl( \delta T_b
 - \cQ \Phi_b \, e^{\varphi} \Bigr) \,,
\ee
with $C$ an integration constant. The corresponding Einstein-frame potential is
\be
 V_{\EF}(\varphi) = e^{2(\varphi-\varphi_\star)} \, V_{\JF}
 = C e^{\varphi-2\varphi_\star} + \frac12 \sum_b \Bigl( \delta T_b e^{2(\varphi-\varphi_\star)}
 - \cQ \Phi_b e^{3\varphi-2\varphi_\star} \Bigr) \,.
\ee
The integration constant $C$ is set by demanding that $V_\EF'$ vanishes at $\varphi=\varphi_\star$:
\be
 C e^{-\varphi_\star}=  \sum_b \left( - \delta T_b + \frac32 \, \cQ \Phi_b \, e^{\varphi_\star} \right) \,,
\ee
leading to the full Einstein-frame potential
\be \label{appFEFpot}
 V_{\EF} = \(\frac12 \sum_b \delta T_b \) \Bigl( e^{2 \psi}
 - 2 \, e^{\psi} \Bigr)
 + \( \frac12 \sum_b \cQ\Phi_b e^{\varphi_\star} \)
 \Bigl( 3 \, e^{\psi} - e^{3\psi} \Bigr) \,,
\ee
where $\psi := \varphi - \varphi_\star$.

The 4D on-brane curvature obtained from the full 6D field equations agrees (by construction) with the curvature obtained from the 4D Einstein equations with $V_\EF$ evaluated at $\varphi_\star$. Using ~\pref{appFEFpot}, we find
\be
 V_{\star} := V_{\EF}(\varphi_\star)
 = \frac12 \sum_b \Bigl( - \delta T_b
 + 2 \, \cQ\Phi_b \, e^{\varphi_\star} \Bigr)  \,,
\ee
which vanishes by virtue of the stabilization condition defining $\varphi_\star$, eq.~\pref{stabilization2}. Notice that this condition is also equivalent to the linearized version of the warped flux-quantization condition, eq.~\pref{warpedquantization}
\ba \label{appwarpedquantization}
 0 &=& \delta \left[ \frac{(\alpha_+ \alpha_-)^{1/2}}{g_\ssR}
 + \frac{1}{2 \pi} \left( \frac{\Phi_o}{\cW^2_o}
 + \frac{\Phi_f}{\cW^2_f} \right) e^{\varphi_\star} \right] \nn\\
 &=& \frac{1}{2 g_\ssR} \sum_b \delta \alpha_b
 + \frac{1}{2 \pi} \sum_b \delta \Phi_b \, e^{\varphi_\star}  \nn\\
 &=& \frac{\kappa^2}{4 \pi g_\ssR} \sum_b \Bigl( \delta T_b - 2 \, \cQ
 \Phi_b \, e^{\varphi_\star} \Bigr) \,,
\ea
which uses $\delta \alpha_b = \kappa^2 \delta L_b/2\pi = \kappa^2 (\delta T_b - \cQ \Phi_b)/2\pi$, as well as the unperturbed rugby-ball relation $\cQ = 2 g_\ssR/\kappa^2$.

\subsection*{Bulk loop corrections to the brane action}
\label{app:4dpotential2}

In general, bulk loops induce a $\phi$-dependence to the brane action, and so generate a nonzero curvature for the on-brane directions. The most UV-sensitive contributions come when very heavy bulk particles circulate in the loop, and because these involve only very short wavelengths they generate local corrections to the brane action. We now argue that these UV-sensitive bulk loops contribute only to $V_\star$ at order $m_\KK^4$, where $m_\KK \simeq 1/r \simeq \cV_2^{-1/2}$ is the Kaluza-Klein scale.

Loops involving comparatively long-wavelength states at the KK scale need not generate only local effects on the branes, but also only give rise to contributions to the low-energy vacuum energy that are of order $\delta V_\star \sim m_\KK^4$ (and so are not larger than the UV loops we examine below). Because $e^{2\phi}$ is the loop-counting parameter in the bulk, an estimate for the size of the UV loop-generated curvature can be found by repeating the above arguments, but now writing $T_b$ and $\Phi_b$ as a series in powers of $e^{2\phi}$, rather than taking them to be $\phi$-independent.

To this end we write
\ba
 \delta T_b &=& \delta T_b^{(0)} + \delta T_b^{(1)} e^{2\phi} + \cdots \nn\\
 \Phi_b &=& \Phi_b^{(0)} + \Phi_b^{(1)} \, e^{2\phi} + \cdots \,,
\ea
where $T_b^{(1)}$ and $T_b^{(0)}$ are $\phi$-independent and similar in size, as are $\Phi_b^{(1)}$ and $\Phi_b^{(0)}$. With these choices we have
\ba
 \delta L_b &=& \delta T_b^{(0)} - \cQ \Phi_b^{(0)} \, e^{\varphi_0}
 + \delta T_b^{(1)} \, e^{2\varphi_0}
 - \cQ \Phi_b^{(1)} \, e^{3\varphi_0} \nn\\
 \hbox{and so} \quad
 \delta L_b' &=& 2 \,\delta T_b^{(1)} \, e^{2\varphi_0}
 -2 \cQ\Phi_b^{(1)} \, e^{3\varphi_0} \,.
\ea

The condition defining the stationary point, $\varphi_\star$, is given by
\ba \label{stabilizationloop}
 0&=&\sum_b\(\delta L_b+\frac12 \, \delta L_b'-\cQ\Phi_b\)\nn\\
 &=& \sum_b \Bigl( \delta T_b^{(0)}- 2 \cQ\Phi_b^{(0)} \,  e^{\varphi_\star}
 +2 \, \delta T_b^{(1)} \, e^{2\varphi_\star}
 -3\cQ\Phi_b^{(1)} \, e^{3\varphi_\star} \Bigr) \,.
\ea
At lowest order this is solved by $\varphi_\star^{(0)}$ satisfying eq.~\pref{appzeroethorder}, and to next-to-leading order the correction, $\delta e^{\varphi_\star}$, satisfies
\be \label{stabilizationloop2}
  \delta e^{\varphi_\star}  \sum_b  2 \cQ\Phi_b^{(0)}
 = \sum_b \Bigl(
 2 \, \delta T_b^{(1)} \, e^{2\varphi_\star^{(0)}}
 -3\cQ\Phi_b^{(1)} \, e^{3\varphi_\star^{(0)}} \Bigr) \,.
\ee
If $\delta T_b^{(1)}$ and $\cQ \Phi_b^{(1)}$ are similar in size then only the first term on the right-hand-side of this last expression dominates. We keep both here because our interest in the main text is in the case where $\delta T_b^{(1)}$ and $\cQ \Phi_b^{(1)} \, e^{\varphi_\star^{(0)}}$ are similar in size.

The corrected Jordan-frame potential due to the brane perturbations then solves eq.~\pref{JFVDE}, or
\be
 \( e^\varphi V_{\JF}\)' = e^{\varphi}  \sum_b \left[
 \frac12 \, \delta T_b^{(0)} - \cQ\Phi_b^{(0)} \, e^{\varphi}
 +2 \, \delta T_b^{(1)} \, e^{2\varphi}
 -\frac52 \, \cQ\Phi_b^{(1)} \, e^{3\varphi} \right] \,,
\ee
which integrates to
\be
 V_{\JF} = C e^{-\varphi} + \sum_b \left[ \frac12\, \delta T_b^{(0)}
 - \frac12 \, \cQ\Phi_b^{(0)} \, e^{\varphi}
 + \frac23 \, \delta T_b^{(1)} \, e^{2\varphi}
 -\frac58 \, \cQ\Phi_b^{(1)} \, e^{3\varphi} \right]\,,
\ee
with $C$ an integration constant, as before.

The Einstein frame potential, $V_{\EF} = e^{2(\varphi-\varphi_\star)} V_{\JF}$, similarly is
\ba
 V_{\EF} &=& C e^{\varphi-2\varphi_\star}
 + \sum_b \left[ \frac12 \, \delta T_b^{(0)} \, e^{2(\varphi-\varphi_\star)}
 - \frac12 \, \cQ\Phi_b^{(0)} \, e^{3\varphi-2\varphi_\star} \right.\nn\\
 &&\qquad\qquad \qquad\qquad \left. + \frac23 \, \delta T_b^{(1)}
 \, e^{4\varphi-2\varphi_\star}
 - \frac58 \, \cQ\Phi_b^{(1)} \, e^{5\varphi-2\varphi_\star} \right] \,.
\ea
As before, enforcing $V_\EF'(\varphi_\star) = 0$ fixes $C$, giving
\be
 Ce^{-\varphi_\star} = \sum_b \left[ - \delta T_b^{(0)}
 + \frac32 \, \cQ\Phi_b^{(0)} \, e^{\varphi_\star}
 - \frac83 \, \delta T_b^{(1)} \, e^{2\varphi_\star}
 +\frac{25}8 \, \cQ\Phi_b^{(1)} \, e^{3\varphi_\star} \right]\,.
\ee
The full next-to-leading Einstein-frame potential then is
\ba
 V_{\EF} &=& \( \frac12 \sum_b \delta T_b^{(0)} \)
 \Bigl( e^{2\psi} - 2\, e^{\psi} \Bigr)
 + \( \frac12 \, e^{\varphi_\star} \sum_b  \cQ\Phi_b^{(0)} \)
 \Bigl( 3\, e^{\psi} - e^{3\psi} \Bigr) \\
 && \qquad + \( \frac23  \, e^{2\varphi_\star}\sum_b \delta T_b^{(1)}\)
 \Bigl( e^{4\psi} - 4 \, e^\psi \Bigr)
 + \( \frac58 \, e^{3\varphi_\star} \sum_b \cQ \Phi_b^{(1)} \)
 \Bigl( 5 \, e^{\psi} - e^{5\psi} \Bigr) \,, \nn
\ea
with $\psi = \varphi - \varphi_\star$. Evaluating this at $\varphi_\star$, and using ~\pref{stabilizationloop}, we find
\be
 V_\star = V_{\EF}(\varphi_\star) = - e^{2\varphi_\star}
 \sum_b \delta T_b^{(1)} +  e^{3\varphi_\star} \sum_b \cQ\Phi_b^{(1)} \,.
\ee
This is clearly of order $e^{2 \varphi_\star}$ if $\delta T_b^{(0)}$, $\delta T_b^{(1)}$, $\Phi_b^{(0)} e^{\varphi_\star}$ and $\Phi_b^{(1)} e^{\varphi_\star}$ are all of the same order. Recalling that the flux-quantization condition relates $\varphi_\star$ to the stabilized extra-dimensional radius by $r_\star$ by $e^{\varphi_\star} \simeq \cO(1/r_\star^2)$, we see that the loop-corrected brane action gives a result of order $1/r_\star^4$, which is also similar to the size of a generic bulk Casimir energy.

\section{No-go results}

There are a number of famous no-go results, that superficially appear to contradict our results. In this appendix we describe two of these, describing why they do not represent real obstructions.

\subsection{Weinberg's no-go theorem}

The best-known obstruction to finding a relaxation mechanism that sets the cosmological constant to zero is due to Weinberg \cite{Wbgcc}. His is a general objection to using scale invariance to solve the cosmological constant problem.
Although his argument is phrased quite generally, it is easier to describe the issues within a simple toy model.

\subsubsection*{Why at first sight scale invariance seems to help}

At first sight, scale invariance provides a very attractive way to approach why the vacuum energy might be zero. To see why, consider the following simple scale-invariant toy theory:
\be
 S = - \int \exd^4x \; \sqrt{-g} \; \left( \frac12 \, \partial_\mu \chi \, \partial^\mu \chi + \overline \psi \gamma^\mu \partial_\mu \psi  + \lambda \, \chi^4 + g \overline \psi \psi \chi \right) \,.
\ee
This action is invariant under the rigid rescalings $\psi \to \zeta^{-3/2} \psi$, $\chi \to \zeta^{-1} \chi$ together with $g_{\mu\nu} \to \zeta^2 g_{\mu\nu}$, although this symmetry is anomalous and so does not survive quantization. However at the classical level it restricts the potential to only have a quartic term, and ensures the scale-invariant point, $\chi = 0$, is a solution to $V'(\chi = 0) = 0$. Because scale invariance precludes the existence of any dimensionful parameters, it also guarantees that the potential vanishes at this scale-invariant minimum: $V(\chi = 0) = 0$.

Suppose we put aside (for now) the anomaly in scale invariance, and ask whether the fact that $V' = V = 0$ is automatically satisfied means that scale invariance can help solve the cosmological constant problem. At first sight the answer is `no', because having $V = 0$ when $\chi = 0$ is not in itself sufficient. It is insufficient because not only does scale invariance ensure $V = 0$; it makes {\em all} masses zero. After all, the cosmological constant problem is the puzzle of why the effective scalar potential is minimized at a value that is much smaller than the other nonzero masses in the problem.

A more promising attempt might be to consider the case where $\lambda = 0$. In this case all values of $\chi$ are equally good as vacua, and for all of these except $\chi = 0$ the mass of the fermion $\psi$ is nonzero, $m = g \chi$, because the scale invariance is spontaneously broken. Since it is broken masses can be nonzero, but notice that the potential energy is nevertheless still minimized (trivially, since $V = 0$) at zero. Scale invariance guarantees that $V=0$ remains true even once scale invariance is broken, because all values of $\chi$ are related to one another by a symmetry (scale transformations), and so $V$ must have the same value for all of them (and so must in particular vanish, because $V=0$ for the scale-invariant point where $\chi = 0$).

Phrased this way, spontaneously broken scale invariance sounds like a promising approach to having vanishing vacuum energy while still having nonzero masses.

\subsubsection*{Weinberg's objection}

Weinberg's objection to the above argument is that, although promising, scale-invariance in itself cannot solve the cosmological constant problem, even assuming that it could be made not anomalous. That is because scale invariance can never preclude quantum corrections from generating a nonzero scalar potential, like $\lambda \, \chi^4$ which we've seen is completely scale invariant. And if such a potential is generated, the only minimum is again the scale invariant point, $\chi = 0$, for which all masses vanish.

The problem with scale invariance is not that quantum corrections raise the minimum of the potential from $V=0$; it is that quantum corrections generically lift the flat direction and make the scale-invariant point the only minimum.

\subsubsection*{Relevance to the 6D model}

Weinberg's analysis is not specific to four dimensions, and applies equally well to extra-dimensional theories that are scale invariant. It is particularly pertinent for the supergravity models discussed in the main text, for which the bulk enjoys a classical scaling symmetry. Although the analog of the scale-invariant point may seem less clear in the extra-dimensional model, it is $\chi = e^{\varphi_0}$ that plays the role described above, since this transforming multiplicatively under a scale transformation rather than shifting. This shows that having a potential minimized only at the scale invariant point corresponds in the extra-dimensional model to having a runaway potential that is only minimized for infinitely large values of the dilaton, $\phi$.

And in the main text we've also seen that in the special case where the branes couple to $\phi$ in the scale-invariant way, the generic form for the classical low-energy potential is $V_{\rm eff} = A \, e^{2 \varphi_0}$, revealing the generic runaway Weinberg's argument requires.

But nothing in this argument precludes finding the minima obtained in the main text. For more general kinds of brane-dilaton couplings the shape of the potential is more complicated since its form is no longer dictated by scale invariance. Nothing forces it to be minimized only at the scale invariant point in this case.

Furthermore, nothing in the argument says {\em how large} the corrections to the potential have to be. In the 6D model described above, supersymmetry in the bulk generically acts to suppress the size of quantum corrections, regardless of whether or not these are scale invariant.

So this argument, while true, doesn't preclude the behaviour found in the main text.

\subsection{What is the 4D mechanism?}

Another general objection to the kind of calculation presented here asks what the ultimate mechanism looks like in 4 dimensions. That is, even though the full theory is extra-dimensional, why can't I ask what the perspective is of a 4D brane-localized observer? After all, if there is a mechanism at work in 4D, this could be more widely useful than a particular higher-dimensional example.

The basic response to this question is that the underlying mechanism at work in the brane back-reaction is higher-dimensional, and cannot be simply seen in a purely 4D framework involving only a small number of 4D fields. Ultimately, this is why the KK scale must be as low as sub-eV energies in order to be relevant to the observed Dark Energy density: if it were higher the extra dimensions could have been integrated out and we would be back to the unsolved problem of understanding why the Dark Energy is small in four dimensions.

Of course the world {\em does} appear four-dimensional below the KK scale, and in this energy range a 4D observer must be able to understand what is going on. But when the KK scale is as low as the Dark Energy scale, $\rho \simeq m_\KK^4$, there really is also no cosmological constant problem in 4D since $\rho$ is as big as the largest UV scale --- {\em i.e.} $m_\KK$ --- would suggest it should be. The essence of the SLED mechanism is that above the KK scale the gravitational response of the vacuum {\em must} be understood in 6D, even though all non-gravitational physics remains 4D (because it is localized on the brane).

\subsubsection*{Arguments why a 4D mechanism is necessary}

But a more subtle objection\footnote{We thank Nima Arkani-Hamed for making this argument to us.} asks why a thought experiment cannot be performed that allows the vacuum energy to be understood within a 4D effective theory, even if the cosmological constant were larger than $m_\KK$. After all, one can imagine adiabatically changing the underlying parameters of the model in such a way as to generate an effective 4D cosmological constant, $\cL_{\rm eff} = -  \sqrt{-g} \; A$, with $A > m_\KK^4$. If so, because energy cannot be directly extracted from $\rho$, no consistency issue would preclude us from analyzing the theory in an effective 4D approximation, provided the Hubble scale remains small enough: $H^2 \simeq A/M_p^2 \ll m_\KK^2$. (If $H$ were to become larger than $m_\KK$ then sufficient energy could be extracted from the time-dependent geometry to excite KK modes and force us outside of the domain of the effective 4D description.)

In this picture, it seems we are again forced to be able to understand what keeps $\rho$ from being large purely within a 4D context.

But again it is the scale invariance that saves the day. As you adiabatically manipulate the underlying parameters in the 6D theory, what is generated is a potential, $V(\varphi_0)$, for the entire flat direction rather than just a cosmological constant. Since the flat direction partially involves the extra-dimensional metric, general covariance precludes generating just a $\varphi_0$-independent constant.

So instead of getting a constant like $\cL_{\rm eff} = - \sqrt{-g} \; A$ one instead gets a potential like $\cL_{\rm eff} = - \sqrt{-g}\; A \, e^{a \varphi_0}$, where $a$ is order unity. But if $A$ rises above $m_\KK^4$, then not only does $V$ rise above $m_\KK^4$, but also so does its derivative, $V'$. Once this is true a 4D description is no longer possible, because the equation of motion for $\varphi_0$ implies that having $V'$ this large generates a time derivative $\dot\varphi_0$ that is equally large, which provides an energy source that can generate KK modes.

The upshot is that there is no effective 4D understanding of the cosmological constant problem; but this does not mean that no solution exists, it simply means that the KK scale cannot be much larger than the observed Dark Energy density. It also means that the existence of a light field, $\varphi_0$, in the low-energy theory is a crucial part of the story, making it unavoidable that there be a scalar-tensor gravity in the long-wavelength limit.

\end{document}